\documentclass[useAMS,usenatbib]{mn2e}

\usepackage {graphicx}
\usepackage {times}

\title[AKARI IR LFs of the local Universe]
{Luminosity Functions of Local Infrared Galaxies with the AKARI:
Implications to the Cosmic Star Formation History and AGN Evolution 
}
\author[Goto et al.]{Tomotsugu Goto$^{1,2}$, 
 \thanks{E-mail:tomo@ifa.hawaii.edu} 
Stephane Arnouts$^3$,
Hanae Inami$^{4}$,
Hideo Matsuhara$^{5}$,
Chris Pearson$^{6}$,
\newauthor Tsutomu T. Takeuchi$^{7}$,
Emeric Le Floc'h$^{8}$,
Toshinobu Takagi,
Takehiko Wada,
\newauthor 
Takao Nakagawa$^{5}$,
Shinki Oyabu$^{2}$,
Daisuke Ishihara$^{7}$,
Hyung Mok Lee$^{9}$,
\newauthor Woong-Seob Jeong$^{10}$,
Chisato Yamauchi$^{5}$,
 S. Serjeant,
 C. Sedgwick$^{11}$,
and 
Ezequiel Treister$^{1}$
\\
$^{1}$Institute for Astronomy, University of Hawaii, 2680 Woodlawn Drive, Honolulu, HI, 96822, USA\\
$^{2}$Subaru Telescope 650 North A'ohoku Place Hilo, HI 96720, USA\\
$^{3}$Canada France Hawaii Telescope, 65-1238 Mamalahoa Hwy, Kamuela, Hawaii 96743 USA \\
$^{4}$Spitzer Science Center, California Institute of Technology, Pasadena, CA 91125, USA\\
$^{5}$Institute of Space and Astronautical Science, Japan Aerospace Exploration Agency, 	     Sagamihara, Kanagawa 252-5210\\
$^{6}$Rutherford Appleton Laboratory, Chilton, Didcot, Oxfordshire OX11 0QX, UK\\
$^{7}$Institute for Advanced Research, Nagoya University, Furo-cho, Chikusa-ku, Nagoya 464-8601\\
$^{8}$CEA-Saclay, Service d'Astrophysique, France\\
$^{9}$Department of Physics \& Astronomy, FPRD, Seoul National University, Shillim-Dong, Kwanak-Gu, Seoul 151-742, Korea	\\
$^{10}$Korea Astronomy and Space Science Institute 61-1, Hwaam-dong, Yuseong-gu, Daejeon, Republic of Korea 305-348\\
$^{11}$Astrophysics Group, Department of Physics,  The Open University, Milton Keynes, MK7 6AA, UK
}
\begin{document}
\maketitle
\label{firstpage}
\begin{abstract}
  Infrared (IR) luminosity is fundamental to understanding the cosmic star formation history and AGN evolution, since their most intense stages are often obscured by dust. 
 However, local IR luminosity function estimates today are still based on the IRAS survey in the 1980s, with wavelength coverage only up to 100$\mu$m.
  The AKARI IR space telescope performed all sky survey in 6 IR bands (9, 18, 65, 90, 140, and 160$\mu m$) with 3-10 times better sensitivity, covering 
  the crucial far-IR wavelengths across the peak of the dust emission. Combined with a better spatial resolution,  AKARI can much more precisely measure the total infrared luminosity ($L_{TIR}$) of individual galaxies, and thus, the total infrared luminosity density in the local Universe.

 By fitting modern IR SED models, 
 we have re-measured $L_{TIR}$ of the IRAS Revised Bright Galaxy Sample, which is a complete sample of local galaxies with $S_{60\mu m}>5.24Jy$. 

 We present mid-IR monochromatic luminosity ($\nu L_{\nu}$) to $L_{TIR}$ correlations for Spitzer $8\mu m$, AKARI $9\mu$m, IRAS $12\mu m$, WISE $12\mu m$, ISO $15\mu m$, AKARI $18\mu m$, WISE $22\mu m$, and Spitzer $24\mu$m filters. These measures  of $L_{MIR}$ are well 
 correlated with  $L_{TIR}$, with scatter ranging 13-44\%. The best-fit $L_{MIR}$-to-$L_{TIR}$ conversions provide us with estimates of  $L_{TIR}$ using only a single MIR band, in which several deep all sky surveys are becoming available such as AKARI MIR and WISE.

Although we found some overestimates of $L_{TIR}$ by IRAS due to contaminating cirrus/sources, the resulting AKARI IR luminosity function (LF) agrees well with that from the IRAS. We integrate the LF weighted by $L_{TIR}$ to obtain a cosmic IR luminosity density of $\Omega_{TIR}$= (8.5$^{+1.5}_{-2.3})\times 10^{7}$ $L_{\odot}$Mpc$^{-3}$, of which 7$\pm1$\% is produced by LIRGs ($L_{TIR}>10^{11}L_{\odot}$), and only 0.4$\pm$0.1\% is from  ULIRGs ($L_{TIR}>10^{12}L_{\odot}$) in the local Universe, in a stark contrast to high-redshift results.

 We separate the contributions from AGN and star-forming galaxies (SFG). 
 SFG IR LF shows a steep decline at the bright-end.
 Combined with high-redshift results from the AKARI NEP deep survey, these data show a strong evolution of $\Omega_{TIR}^{SF}\propto$(1+z)$^{4.0\pm0.5}$, and  $\Omega_{TIR}^{AGN}\propto$(1+z)$^{4.4\pm0.4}$. For  $\Omega_{TIR}^{AGN}$, the ULIRG contribution exceeds that from LIRG already by z$\sim$1.
 A rapid evolution in both $\Omega_{TIR}^{AGN}$ and $\Omega_{TIR}^{SFG}$ suggests the correlation between star formation and black hole accretion rate continues up to higher redshifts. 
 We compare the evolution of $\Omega_{TIR}^{AGN}$ to that of X-ray luminosity density.
 The $\Omega_{TIR}^{AGN}/\Omega_{X-ray}^{AGN}$ ratio shows a possible increase at $z>1$, suggesting an increase of obscured AGN at $z>1$. 
\end{abstract}

\begin{keywords}
galaxies: evolution, galaxies:interactions, galaxies:starburst, galaxies:peculiar, galaxies:formation
\end{keywords}

\section{Introduction}

 
To understand the cosmic history of star formation and black hole growth, we must understand infrared (IR) emission; the more intense star formation, the more deeply it is embedded in the dust, hence, such star formation is not visible in UV but in the infrared. Similarly, AGN evolutionary scenarios predict that they are heavily obscured at their youngest, Compton-thick stage \citep{2009ApJ...696..110T}. 
 The Spitzer and AKARI satellites revealed large amounts of infrared emission in the high-redshift Universe, showing strong evolution in the infrared luminosity density \citep{2005ApJ...632..169L,2005ApJ...630...82P,2006MNRAS.370.1159B,2007ApJ...660...97C,2009A&A...496...57M}. 
 For example, at z=1, \citet{Goto_NEP_LF} estimated 90\% of star formation activity is hidden by dust.
However, the key baseline of these evolution studies is still at z=0, using a local infrared LF from the IRAS survey in 1980s.

 For more than 25 years, bolometric infrared luminosities ($L_{TIR,8-1000\mu m}$) of local galaxies have been estimated using a simple polynomial presented by  \citet{1987PhDT.......115P}, obtained assuming a simple blackbody and dust emissivity. Furthermore, the reddest filter of IRAS is 100$\mu m$, which does not span the peak of the dust emission for most galaxies, leaving a great deal of uncertainty. 
A number of studies found cold dust that cannot be detected with the IRAS. For example, \citet{2001MNRAS.327..697D} detected such cold dust with $T\sim20$K using SCUBA 450,850 $\mu$m observation. 
\citet{2009MNRAS.397.1728S} detected cold galaxies with SED peaks at longer wavelengths using Spitzer/MIPS. These results cast further doubt on  $L_{TIR}$ estimation made with only $<100\mu$m photometry.  
More precise estimate of local $L_{TIR}$ and thus the local IR luminosity function (LF) have been long awaited, to be better compared with high redshift work. 

AKARI, the Japanese infrared satellite \citep{2007PASJ...59S.369M}, provides the first chance to  rectify the situation since IRAS; AKARI performed an all-sky survey in two mid-infrared (centered on 9 and 18 $\mu m$) and four far-infrared bands (65,90, 140, and 160$\mu m$). Its 140 and 160$\mu m$ are especially important to cover across the peak of the dust emission, allowing us to accurately measure the Rayleigh-Jeans tail of the IR emission. The time is ripe in terms of modeling as well; several sophisticated infrared SED models have become available \citep{2001ApJ...556..562C,2002ApJ...576..159D,2003MNRAS.338..555L,2007A&A...461..445S}, from which we can not only obtain accurate $L_{TIR}$ estimates, but also constrain the  astrophysics.

In this work, we aim to re-measure local $L_{TIR}$, and thereby the IR LF of the Revised Bright Galaxy Sample \citep[RBGS,][]{2003AJ....126.1607S}, which is a complete sample of local IR galaxies. This work provides us an important local benchmark to base future evolution studies at high redshift both by the current AKARI and Spitzer satellites, and by next-generation IR satellites such as Herschel, WISE, JWST, and SPICA.
We adopt a cosmology with $(h,\Omega_m,\Omega_\Lambda) = (0.75,0.3,0.7)$ for a comparison purpose.

\section{The AKARI all sky survey and RBGS}\label{Data}
In this work, we use the $\beta$1 version of the AKARI/IRC bright source catalog and the $\beta$2 version of the AKARI/FIS bright source catalog.
 The 5 $\sigma$ sensitivities in the AKARI IR filters ($S9W,L18W,N60,WS,WL$ and $N160$) are 0.05, 0.09, 2.4, 0.55, 1.4, and 6.3 Jy \citep{IRCpaper,2009AIPC.1158..169Y}.
 In addition to the much improved sensitivity and spatial resolution over its precursor (IRAS), the presence of 140 and 160$\mu m$ bands is crucial to measure the peak of the dust emission in the FIR wavelength, and thus the $L_{TIR}$ of galaxies, especially ones with lower dust temperature.

\begin{figure}
\begin{center}
\includegraphics[scale=0.6]{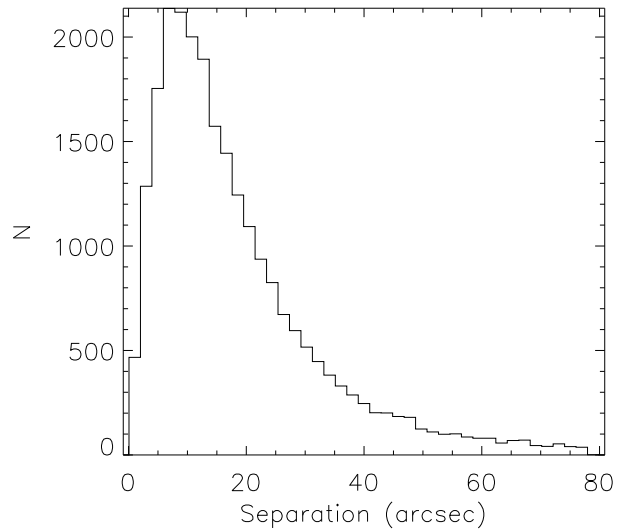}
\end{center}
\caption{
Angular separation between AKARI FIS and IRAS sources. We matched sources within 60 arcsec of separation.
}\label{fig:separation}
\end{figure}

We have cross-correlated the AKARI bright source catalog with the Revised Bright Galaxy Sample \citep[RBGS;][]{2003AJ....126.1607S}, using a matching radius of 60 arcsec, as shown in Fig.\ref{fig:separation}.
The RBGS is a complete, flux-limited survey of all extragalactic objects with 60$\mu m$ flux density greater than 5.24Jy, covering the entire sky at $|b|>5$ deg. All 629 objects in this catalog have measured spectroscopic redshifts either from optical or millimeter/radio (CO or HI data).
The maximum redshift of our sample is 0.087.
 Because of the completeness and availability of the spectroscopic redshifts, this catalog is suitable to construct local infrared luminosity functions (LFs). In addition, each galaxy has a measured IR luminosity from the IRAS survey, allowing us to compare with the AKARI-based one.
Due to the difference in sky coverage (IRAS covers 96\% of the sky, and AKARI covers 94\%), 24 galaxies out of 629 RBGS galaxies did not have an AKARI counterpart. We removed these galaxies from our analysis, but applied corresponding completeness corrections to our statistical analysis. 

One important caveat is that the RBGS is a flux-limited sample from IRAS ($S_{60\mu m}>5.24Jy$). 
Therefore, if the IRAS $S_{60\mu m}$ flux was an overestimate as we discuss later, our work does not address incompleteness in the sample selection.
However, at this very low-redshift (average redshift of 0.0082), the RBGS is still the largest complete sample of IR galaxies with spectroscopic redshift.

\section{Infrared luminosity}
\subsection{Estimating total IR luminosity}\label{sec:LTIR}
  
\begin{figure}
\begin{center}
\includegraphics[scale=0.4]{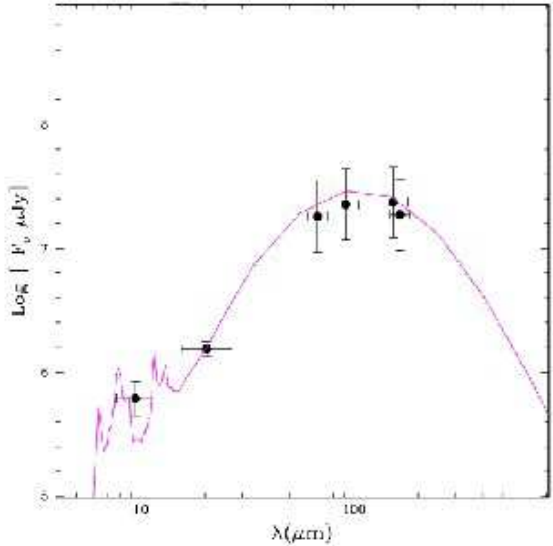}
\end{center}
\caption{
An example of the SED fit. We fit the AKARI 6-band photometry to the SED model of \citet{2001ApJ...556..562C} to estimate  $L_{TIR}$.
}\label{fig:SED_fit}
\end{figure}

For these galaxies, we estimated new total IR luminosities ($L_{TIR}$) based on the AKARI photometry using the {\ttfamily  LePhare} code \footnote{http://www.cfht.hawaii.edu/$^{\sim}$arnouts/lephare.html} to fit the infrared part ($>$7$\mu$m) of the SED and estimate TIR luminosity. 
We fit our AKARI FIR photometry with the SED templates from \citet[CHEL hereafter][]{2001ApJ...556..562C}.
  Although the shapes of these SEDs are luminosity-dependent, the large baseline from AKARI observations  ($S9W,L18W,N60,WS,WL$ and $N160$) allows us to  adopt a free scaling to get
   the best SED fit,  which is then rescaled to derive $L_{TIR}$.  
Since the AKARI catalogs are the $\beta$-version, we adopted a minimum error of 25\%.
 Fig.\ref{fig:SED_fit} shows an example of the SED fit.
 At the median redshift of 0.0082 of the RBGS, peculiar velocity is not negligible. Therefore, instead of the measured redshift, we used the distance, which is corrected for heliocentric redshift using the cosmic attractor model \citep{2000ApJ...529..786M} for the SED fitting and to compute $L_{TIR}$. In this work,  $L_{TIR}$ is measured in the wavelength range of 8-1000$\mu$m.

We chose  not to use IRAS 12,25,60 and 100 $\mu$m fluxes  in the SED fitting. 
Due to the detector pixel size,  AKARI has significantly better spatial resolution (30-40'') than IRAS (1-2').
In a few \% of cases, AKARI measured significantly lower flux than the IRAS, because AKARI can subtract the background cirrus better, and/or can separate nearby confusing sources better. (See \citet{2007PASJ...59S.429J} for more details.)

\begin{figure}
\begin{center}
\includegraphics[scale=0.5]{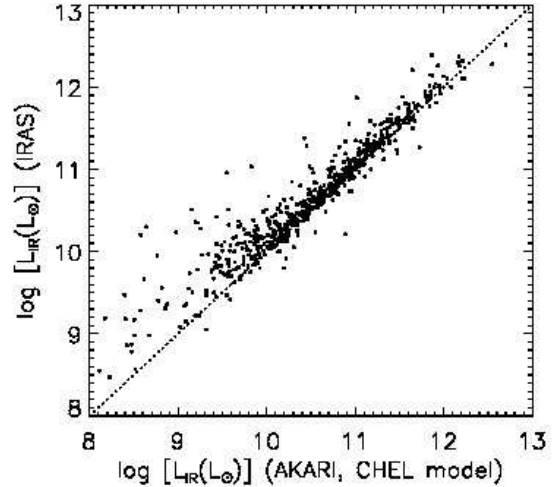}
\end{center}
\caption{
 $L_{TIR}$ measured by the AKARI is compared with those measured by the IRAS (Eq.\ref{eq:sanders}) for the RBGS.
}\label{fig:IRAS_vs_AKARI}
\end{figure}

In Fig.\ref{fig:IRAS_vs_AKARI}, we compare $L_{TIR}$ measured by AKARI with those by IRAS.
The IRAS-based $L_{TIR}$ is measured using the following equation.
\begin{eqnarray}\label{eq:sanders}
  L_{TIR} (L_\odot) &\equiv& 4.93 \times 10^{-22} [ 13.48 L_\nu(12\mu\mbox{m})+5.16L_\nu(25\mu\mbox{m})
    \nonumber \\
    &&+2.58L_\nu(60\mu\mbox{m}) + L_\nu(100\mu\mbox{m})] (erg s^{-1} Hz^{-1}); 
\end{eqnarray}
This equation is obtained by fitting a single temperature dust emissivitly model ($\epsilon \propto \nu^{-1}$) to the flux in all four IRAS bands and should be accurate to $\pm$5\% for dust temperatures in the range of 25-65K \citep{1987PhDT.......115P} .
The $L_{TIR}$ measured by AKARI agrees well with those by IRAS. 
However, the scatter increases toward lower luminosity in $L_{TIR}$ measured by AKARI than by IRAS, especially at log($L_{TIR})<10$. This is primarily because AKARI's higher spatial resolution can separate background cirrus or confusing sources better than IRAS, resulting in lower $L_{TIR}$.


\subsection{Model-to-model variation}\label{sec:model}

\begin{figure*}
\begin{center}
\includegraphics[scale=0.45]{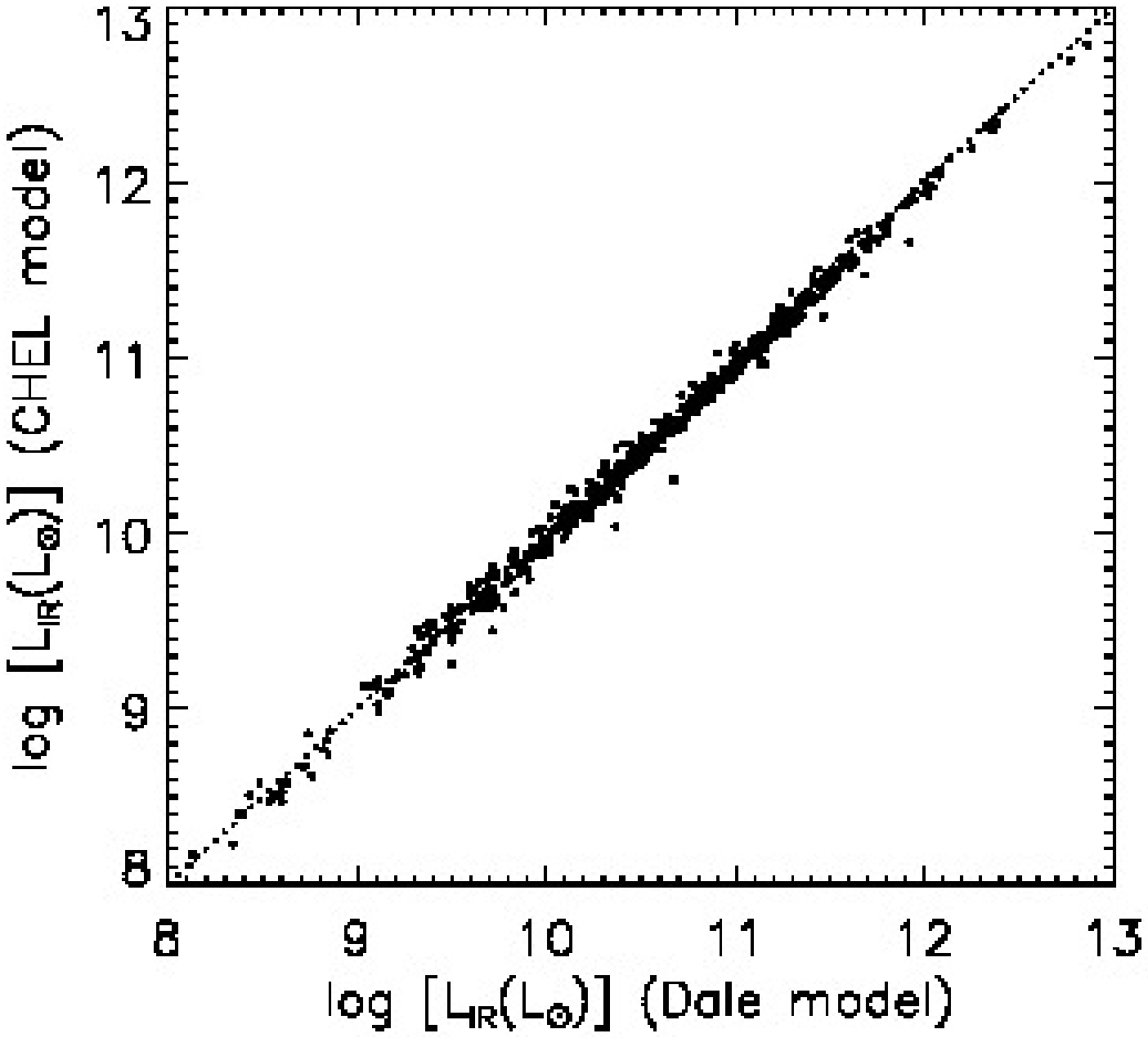}
\includegraphics[scale=0.45]{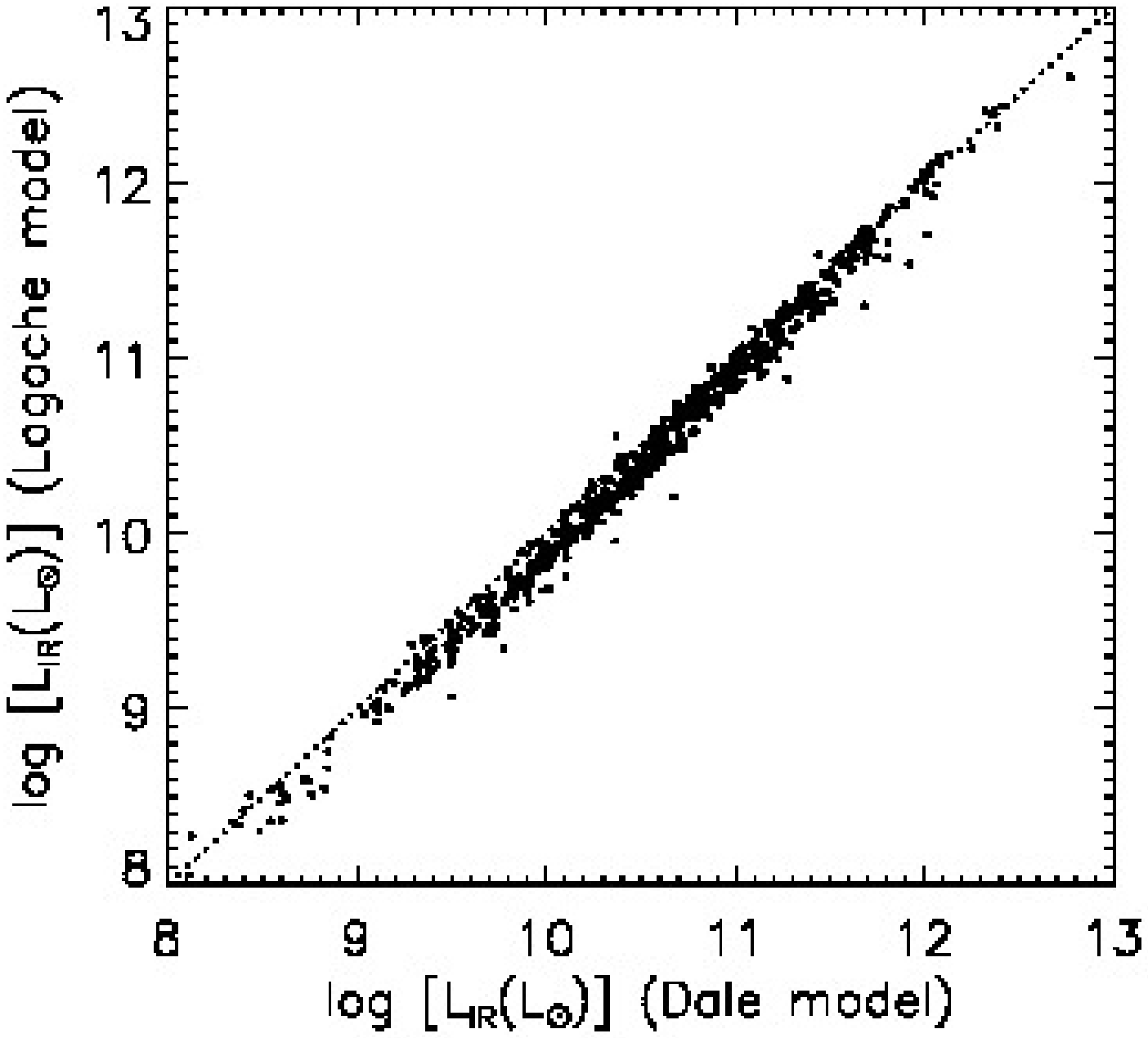}
\includegraphics[scale=0.45]{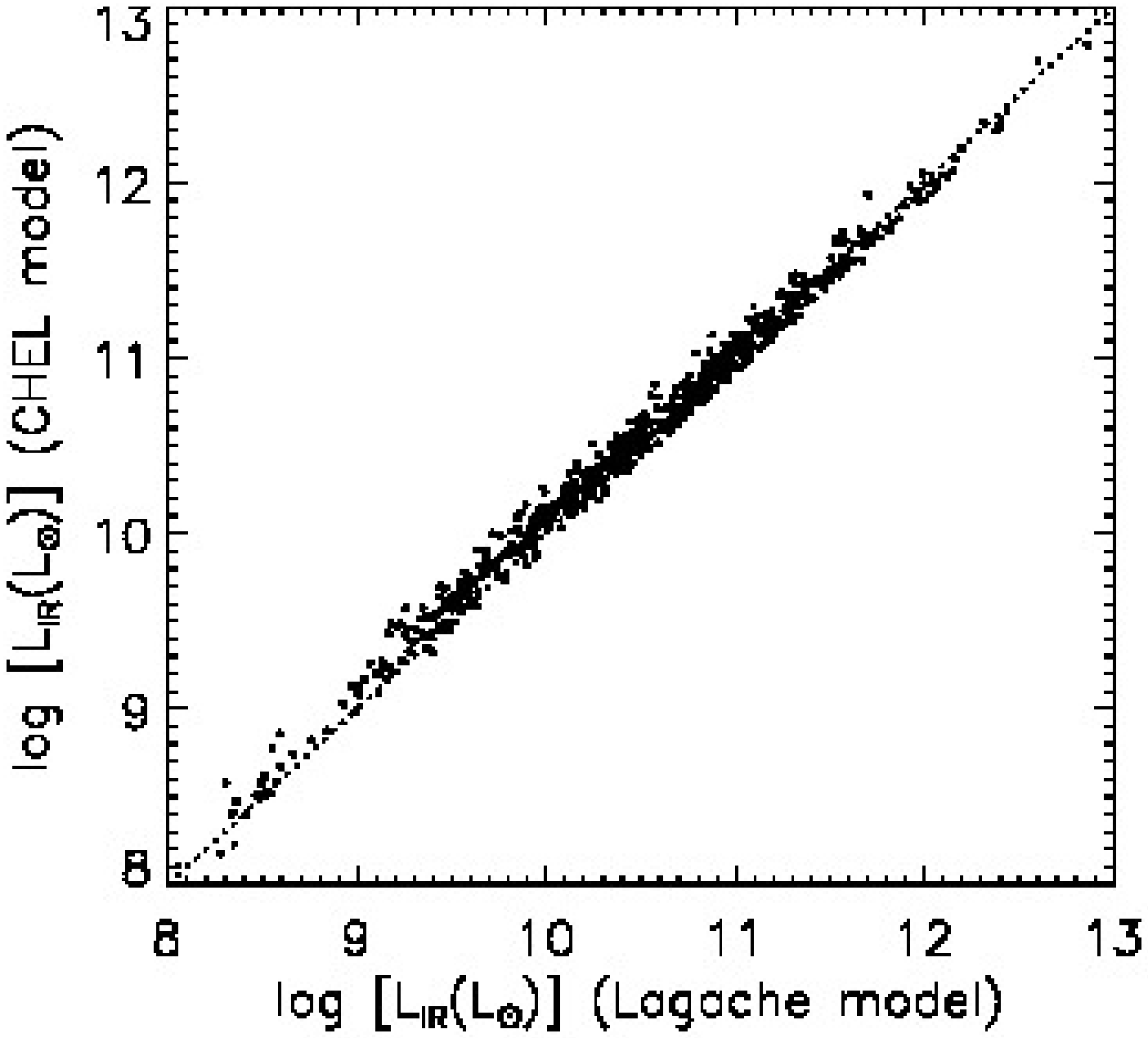}
\end{center}
\caption{
Comparing $L_{TIR}$ measured using three different SED models \citep{2001ApJ...556..562C,2002ApJ...576..159D,2003MNRAS.338..555L}.
 Scatter and offsets between the models are summarized in Table \ref{tab:scatter}.
}\label{fig:L_tir}
\end{figure*}

In this section, we check the SED model dependence of $L_{TIR}$ measurements. 
The IR SED models we used are \citet{2001ApJ...556..562C,2002ApJ...576..159D,2003MNRAS.338..555L}.
We used these models to compute  $L_{TIR}$ exactly as described in Section \ref{sec:LTIR}.
In Fig.\ref{fig:L_tir}, we plot the measured  $L_{TIR}$  against the model results, and find a tight correlation. 
As shown in Table \ref{tab:scatter},  the standard deviations between different $L_{TIR}$s are 10-24\%. 
The agreements are quite good considering  the large flux errors ($\sim 25\%$) associated with infrared photometry.
Median offsets between them are $-13\sim+$24\%, which could also be caused by calibration problems.
 In Table \ref{tab:scatter}, the CHEL model shows smallest offsets from other models. 
 For k-correction and for $L_{TIR}$ estimation (Section \ref{sec:LTIR}), 
 we adopt CHEL models, which  give luminosities in between 
   the Dale models  (11\% higher) and Lagache models  (13\% lower) .

Overall, there are good agreements between models, suggesting that once you have FIR photometry ($WL$ and $N160$ in this work), the model-dependent
uncertainty  in  $L_{TIR}$ is small. This justifies our use of a single SED library for the final $L_{TIR}$ measurement.

\begin{table}
 \centering
  \caption{Comparison of $L_{TIR}$ estimates with different SED models 
 \citep{2001ApJ...556..562C,2002ApJ...576..159D,2003MNRAS.338..555L}.
  Fig.\ref{fig:L_tir} presents corresponding plots.
}\label{tab:scatter}
  \begin{tabular}{@{}lrrlllcccc@{}}
  \hline
   Models & $\sigma$ (\%) & Offset (\%)\\
 \hline
 \hline
CHEL vs IRAS    & 44  & -23\\
Dale vs Lagache    & 24  & 24\\ 
Dale vs CHEL    & 10   & 11\\
Lagache vs CHEL & 22  & -13\\ 
 \hline
\end{tabular}
\end{table}


\subsection{MIR-$L_{TIR}$ correlation}\label{sec:mir_correlation}

A good correlation between Mid-IR monochromatic luminosity and $L_{TIR}$ is known to exist.
This is especially true for SF galaxies because the rest-frame MIR luminosity is dominated by prominent PAH features such as at 6.2, 7.7 and 8.6 $\mu$m, and the $L_{TIR}$ is by dust emission heated by the SF activity; i.e., both of these are good indicators of SF activity.
Observationally MIR detectors are more sensitive than in the FIR. Therefore, if we obtain accurate conversions from MIR to $L_{TIR}$, they will bring $L_{TIR}$ measurement to much more galaxies.

AKARI covers both the MIR and FIR range with its 6 band-passes, measuring both MIR and total luminosity well. 
In addition, because it is an all-sky survey, it provides the 6-band photometry for $\sim$600 RBGS galaxies.
 Here we have one of the best opportunities to investigate the $L_{MIR}$-$L_{TIR}$ correlation.

 We compute the MIR monochromatic luminosity at 8, 9, 12, 15, 18, 22, and 24 $\mu$m using the AKARI photometry at 9 and $18\mu$m, and the best-fit CHEL model from Section \ref{sec:LTIR}. 
For the definitions of monochromatic luminosity, we tried to choose luminosity in passbands of existing IR satellites, for precise and easily-reproducible definition, at the same time avoiding narrow emission/absorption features adding too much noise to the resulting relation. 
 In addition, this choice allows users to apply the relations easily to observed fluxes, at least for low-redshift galaxies whose k-correction is small. 
 We checked this choice of monochromatic filters did not have much effect compared to the exact luminosity at a certain single wavelength.  We later show that MIR-$L_{TIR}$ correlations for $WISE 12\mu m$ and $IRAS 12\mu m$ are almost identical.
 Our choices of filters are Spitzer $8\mu m$, AKARI $9\mu$m, IRAS $12\mu m$, WISE $12\mu m$, ISO $15\mu m$, AKARI $18\mu m$, WISE $22\mu m$, and Spitzer $24\mu$ m.

For $L_{Spitzer 8\mu m}$,  $L_{AKARI 9\mu m}$,  $L_{IRAS 12\mu m}$ and $L_{WISE 12\mu m}$, we started from the observed 9$\mu$m flux, then used the best-fit SED (to the 6 bands) from the CHEL  model ($\S$ \ref{sec:LTIR}) to color-correct the observed AKARI 9$\mu$m flux to $L_{MIR}$ in each filter. 
 To be specific, we computed $L_{Spitzer 8\mu m}$,  $L_{AKARI 9\mu m}$, $L_{IRAS 12\mu m}$ and  $L_{WISE 12\mu m}$ through the filter response function of the Spitzer/IRAC $8\mu m$, the AKARI  $9\mu m$,  the IRAS $12\mu$m, and WISE $12\mu m$ filters. 
Here, the best-fit model is only used to color-correct the observed 9$\mu$m flux, and thus, the obtained $L_{MIR}$ is not an integration of flux of the best-fit model.

\begin{figure*}
\begin{center}
\includegraphics[scale=0.4]{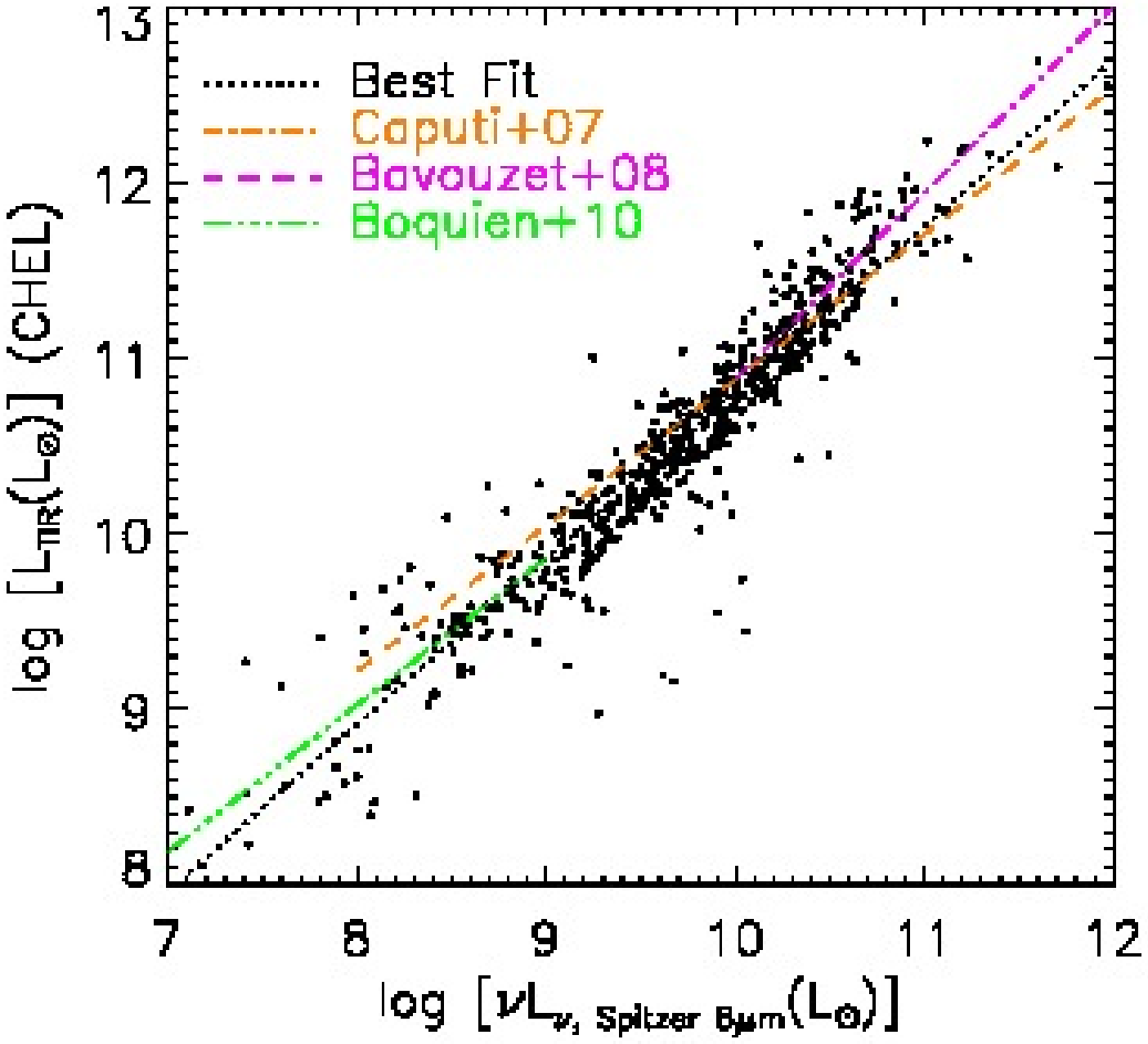}
\includegraphics[scale=0.4]{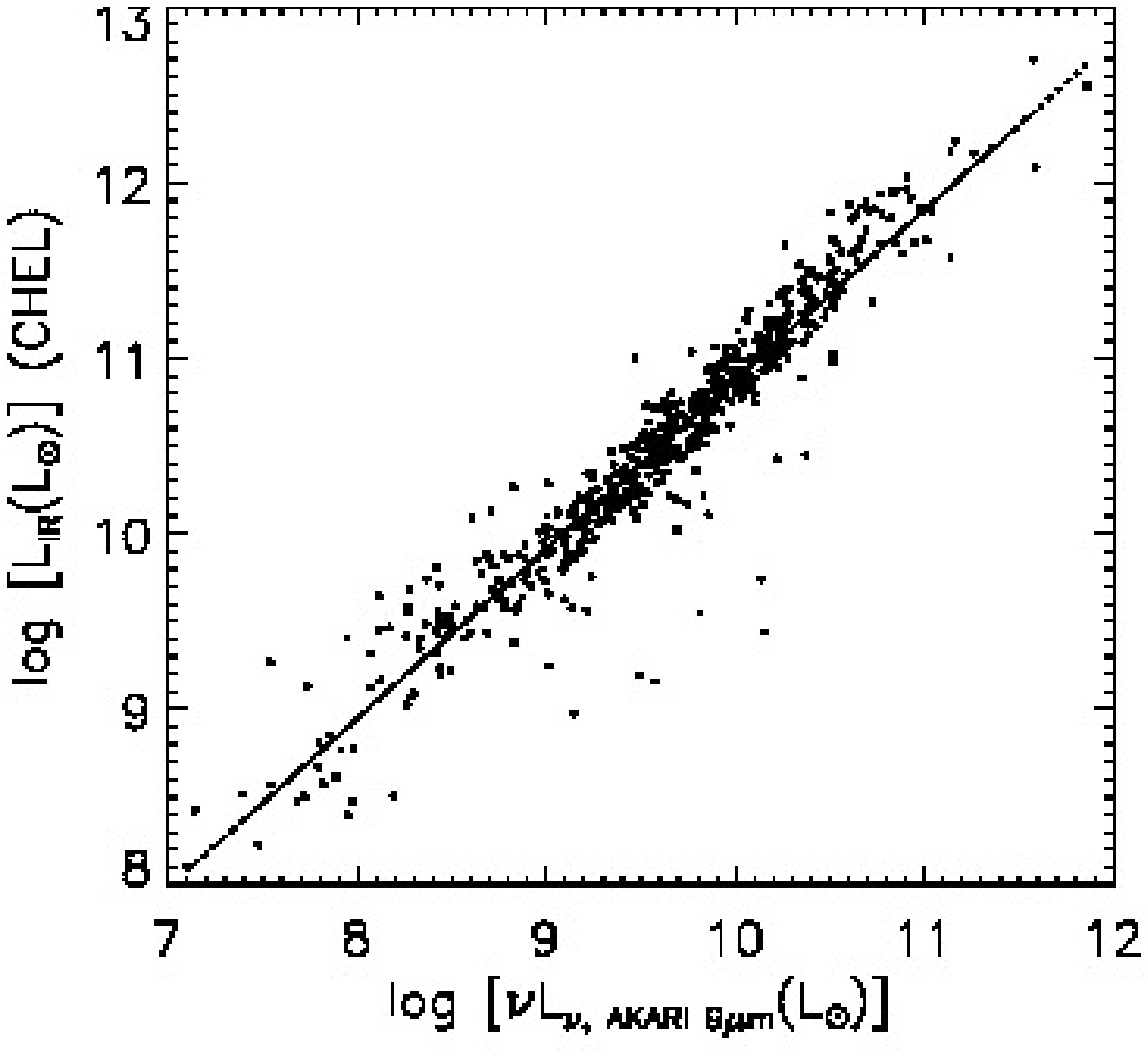}
\includegraphics[scale=0.4]{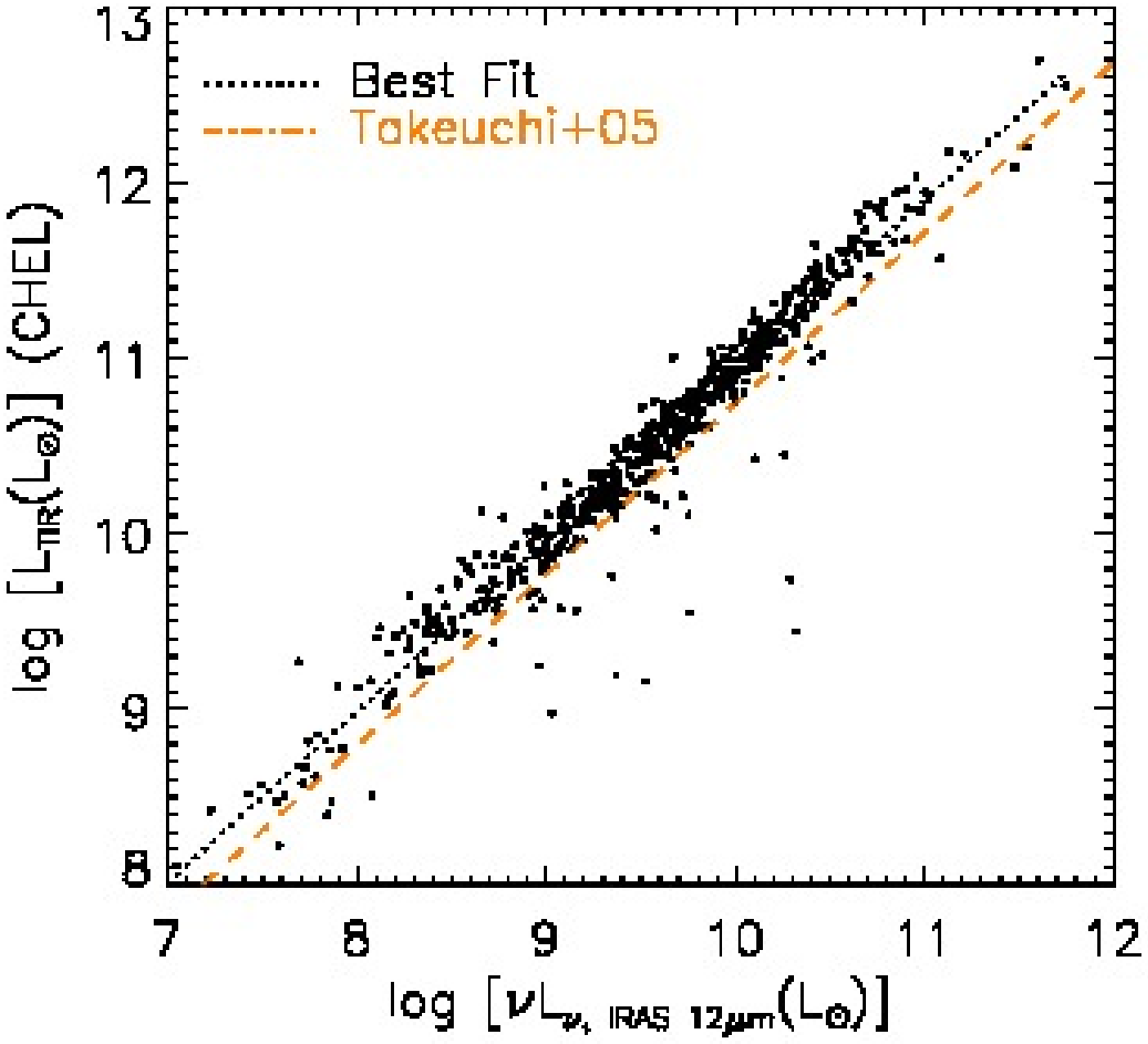}
\includegraphics[scale=0.4]{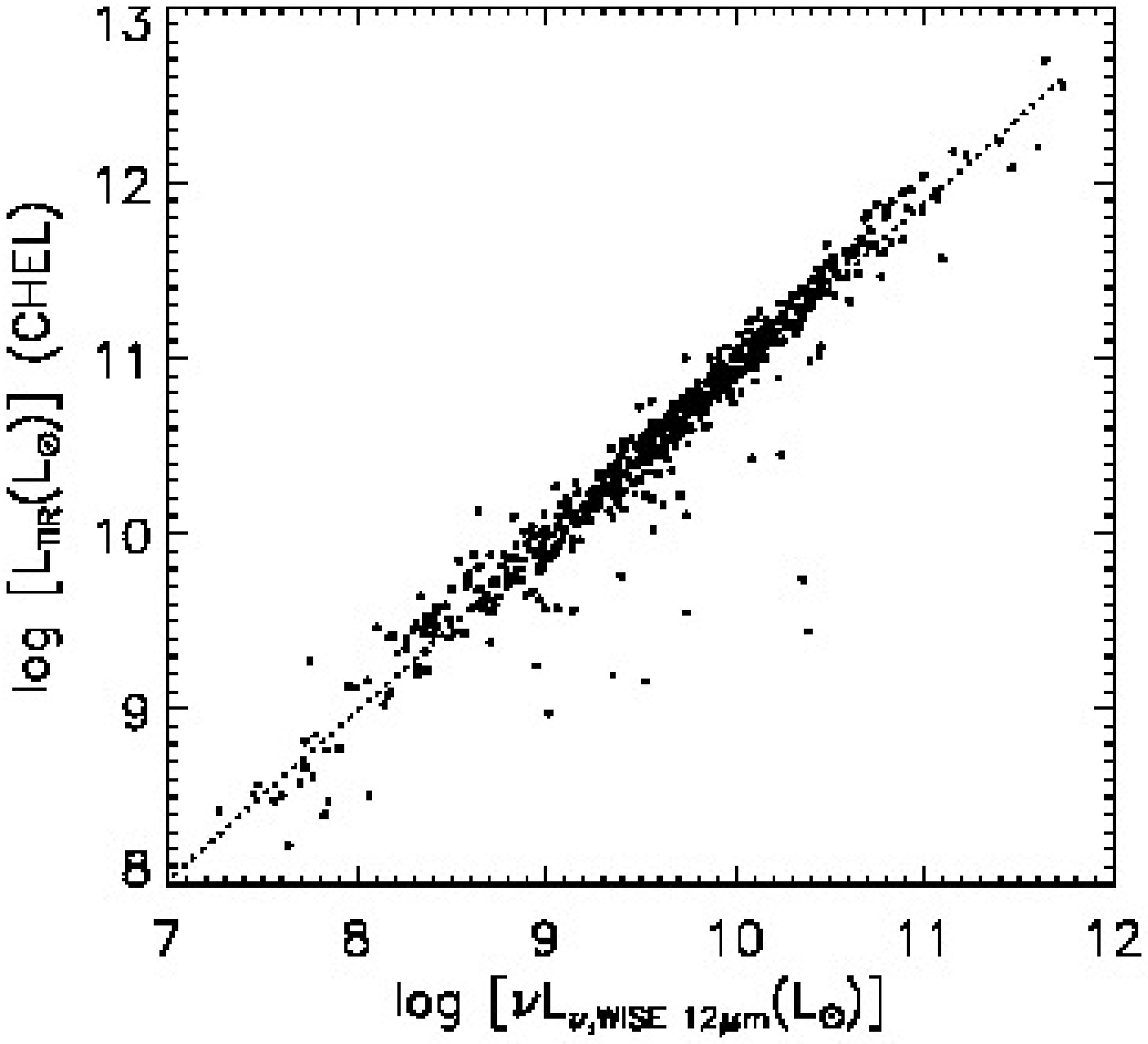}
\includegraphics[scale=0.4]{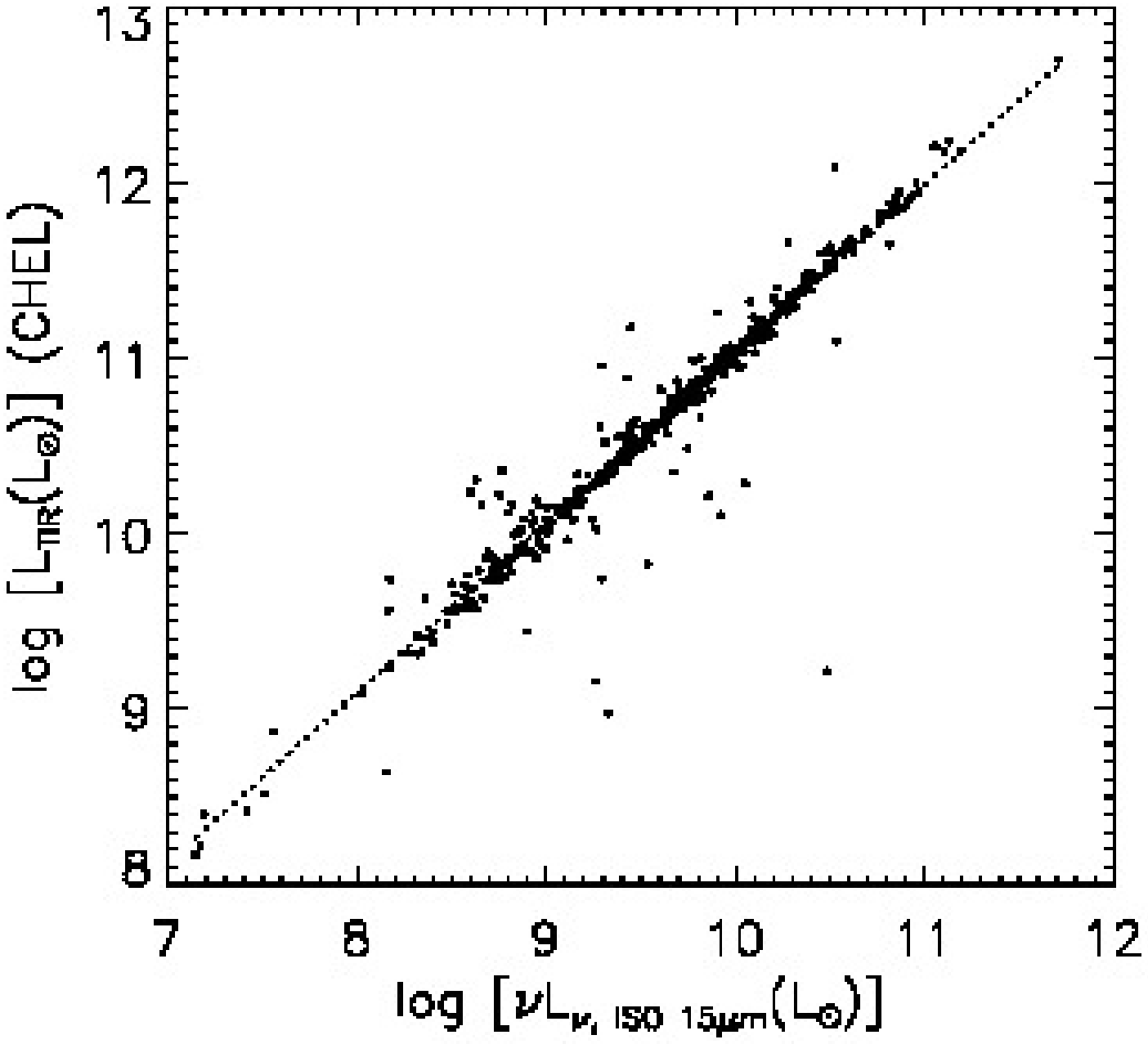}
\includegraphics[scale=0.4]{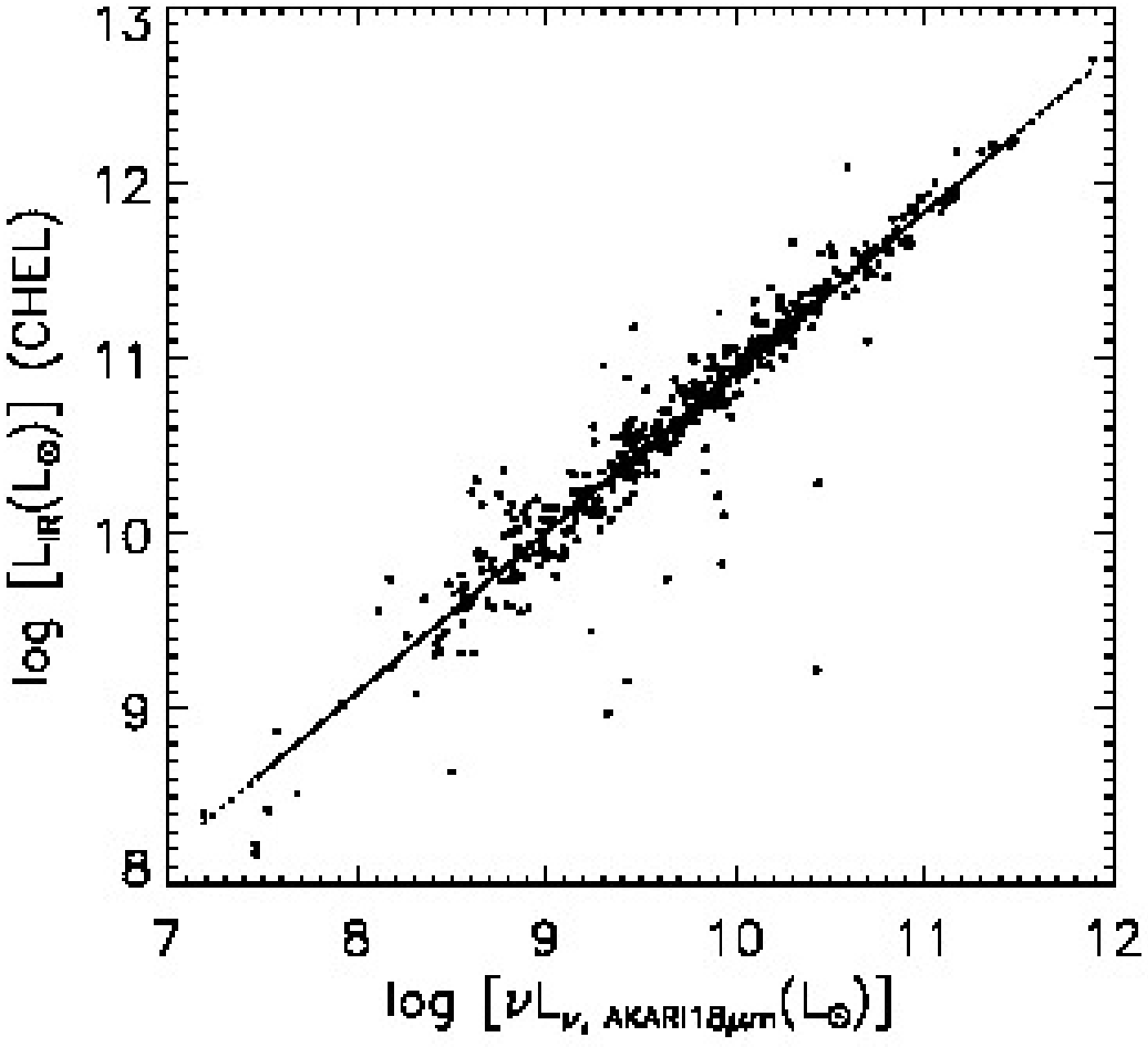}
\includegraphics[scale=0.4]{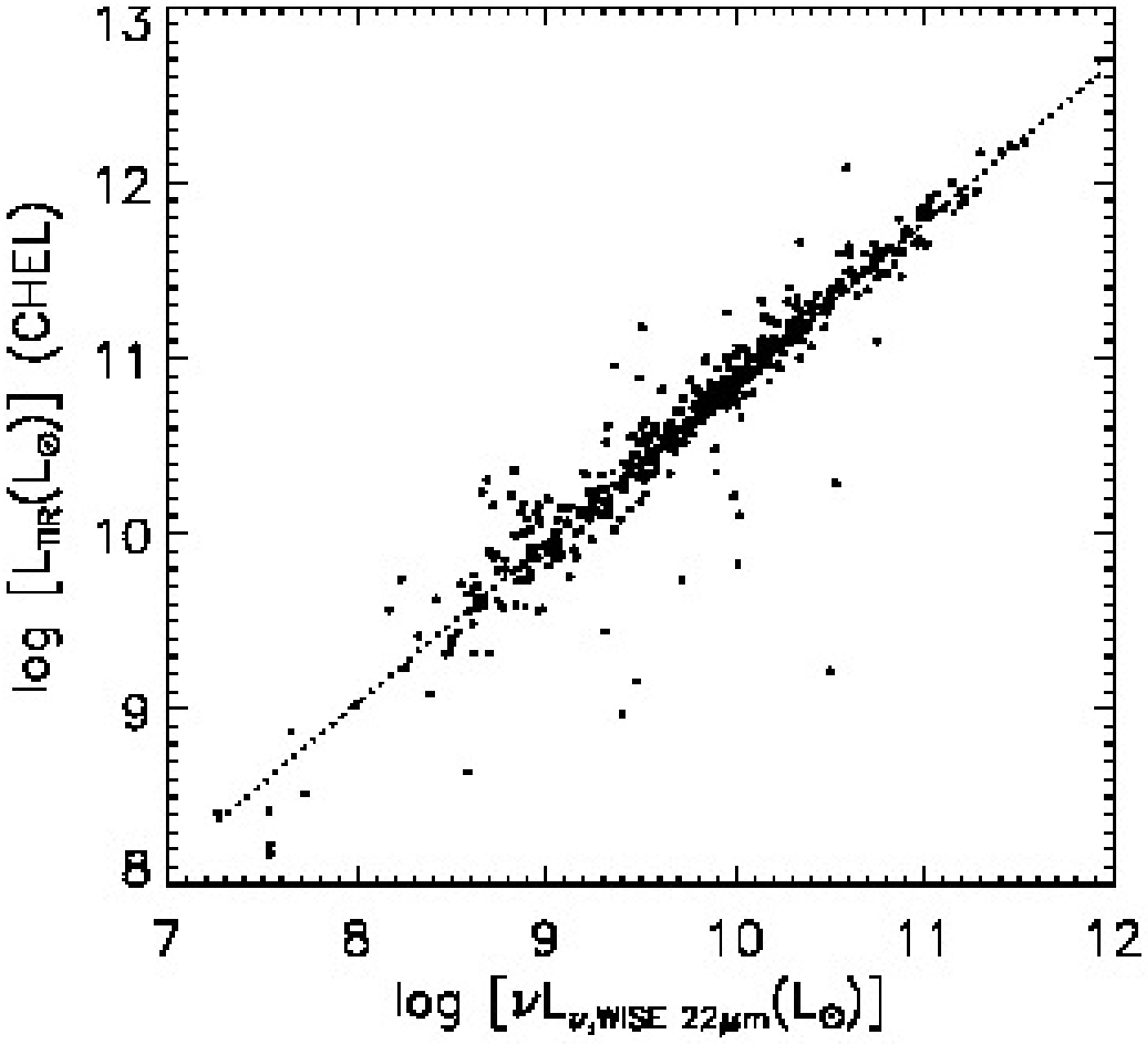}
\includegraphics[scale=0.4]{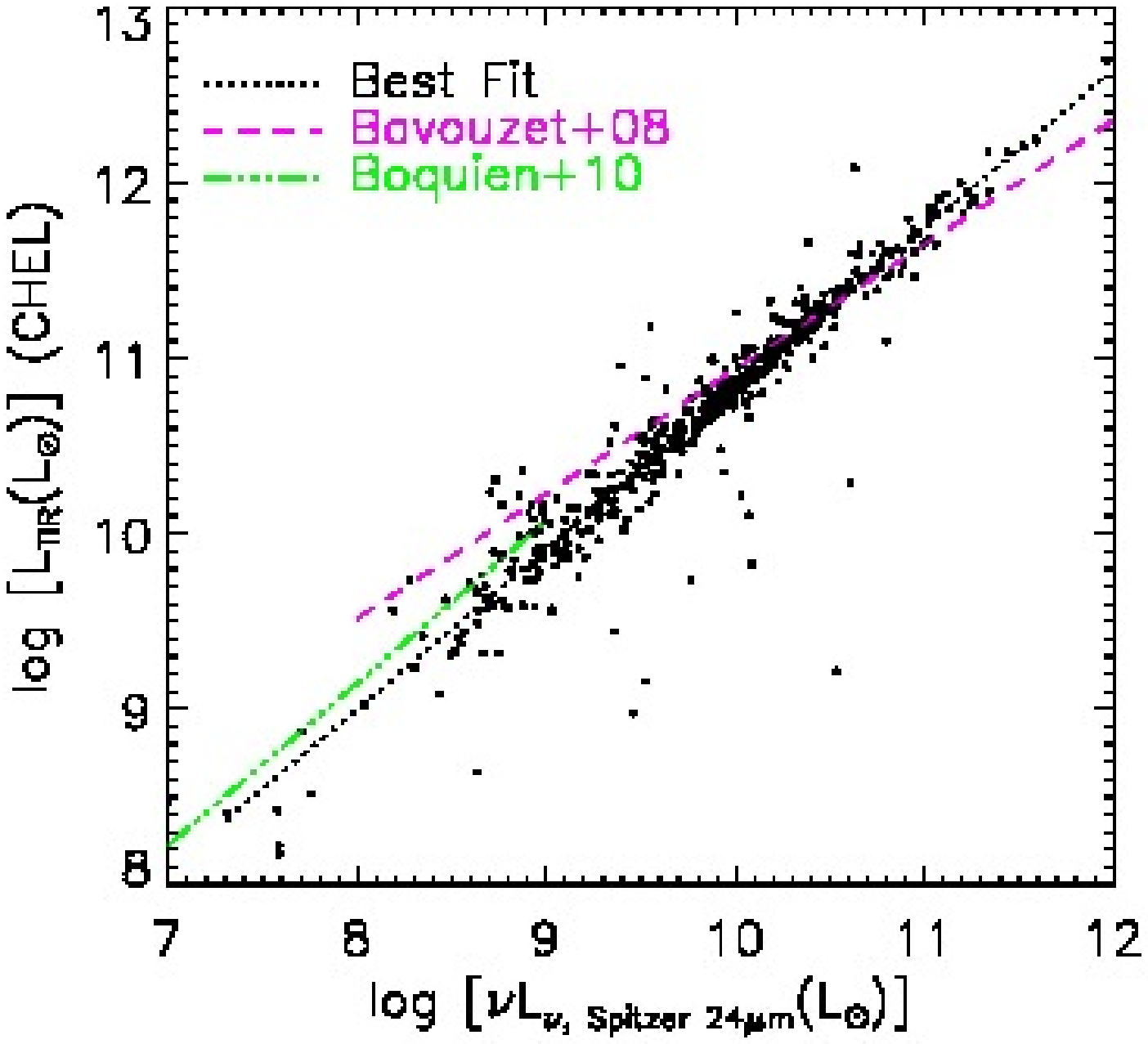}
\end{center}
\caption{
 Correlations between $L_{TIR}$ and monochromatic luminosities (8,9,12,15,18, 22and 24 $\mu$m).
 The lines show the best-fit polynomial, based on which we present conversions.
 When available, conversions from the literature  \citep{2005A&A...432..423T,2007ApJ...660...97C,2008A&A...479...83B,2010ApJ...713..626B} are overplotted for a comparison.
}\label{fig:relation}
\end{figure*}

Similarly, for $L_{ISO 15\mu m}$,  $L_{AKARI 18\mu m}$,  $L_{WISE 22\mu m}$, and $L_{Spitzer 24\mu m}$, we started from the observed AKARI 18$\mu$m flux, color-corrected the flux using the color-correction from the best-fit SED model in $\S$ \ref{sec:LTIR}, 
then converted them to luminosity in each passbands.  
 This color-correction is very small ($\sim$1\% at most) because AKARI's 9 and 18 $\mu m$ are very close to other passbands.
 Thus, the results in Fig.\ref{fig:relation} show almost observed mid-IR luminosity against observed $L_{TIR}$.
 
 In Fig.\ref{fig:relation}, we show monochromatic luminosity in 8,9,12,15,18, 22 and 24 $\mu$m against $L_{TIR}$ measured using the CHEL model. 
 In Fig.\ref{fig:relation}, there exist good correlations in each panel.
 In fact, the correlations look like a simple scaling relation in all panels, i.e., the stronger the MIR-emission, the larger the $L_{TIR}$.
The results suggest indeed monochromatic luminosity in mid-IR range represents the  $L_{TIR}$ well.

  We fit a linear equation to the relation obtaining the following results. Note that the unit is in solar luminosity.

The best-fit relation between  $L_{8\mu m}$ and  $L_{TIR}$ is 
\begin{eqnarray}\label{eq:8um}
  L_{TIR} = (20\pm5) \times \nu L_{\nu,8\mu m}^{0.94\pm0.01} (\pm 44\%).
\end{eqnarray}

The best-fit relation between  $L_{12\mu m}$ and  $L_{TIR}$ is 
\begin{eqnarray}\label{eq:12um}
  L_{TIR} = (17\pm4) \times \nu L_{\nu,12\mu m}^{0.96\pm0.01} (\pm 25\%).
\end{eqnarray}


The best-fit relation between  $L_{15\mu m}$ and  $L_{TIR}$ is 
\begin{eqnarray}\label{eq:15um}
  L_{TIR} = (24\pm8) \times \nu L_{\nu,15\mu m}^{0.96\pm0.01} (\pm 13\%).
\end{eqnarray}

The best-fit relation between  $L_{24\mu m}$ and  $L_{TIR}$ is 
\begin{eqnarray}\label{eq:24um}
  L_{TIR} = (51\pm19) \times \nu L_{\nu,24\mu m}^{0.91\pm0.01} (\pm 24\%).
\end{eqnarray}

The logarithmic slopes of the relations are very similar to each other, perhaps reflecting the fact that most of these IR filters are wide and have some overlaps in wavelength coverage. This also means in this wavelength range (8-24$\mu$m),  and within broad filters, the SED shape does not depend much on $L_{TIR}$.
 However this result is at low redshift, and whether this applies at higher redshift needs to be examined separately. 
We note that  $L_{\nu,15\mu m}$ shows a less dispersed relation with $L_{TIR}$.

For $L_{8\mu m}$, \citet{2007ApJ...660...97C} (eq.\ref{8um_equation_caputi}), \citet{2008A&A...479...83B} (eq.\ref{8um_equation_bavouzet}), and \citet{2010ApJ...713..626B} (eq.\ref{8um_equation_boquien}) present a correlation.
\begin{eqnarray}\label{8um_equation_caputi}
L_{TIR}=1.91\times (\nu L_\nu)^{1.06}_{rest 8\mu m} (\pm 55\%) 
\end{eqnarray}
\begin{eqnarray}\label{8um_equation_bavouzet}
L_{TIR}=377.9\times (\nu L_\nu)^{0.83}_{rest 8\mu m} (\pm37\%) 
\end{eqnarray}
\begin{eqnarray}\label{8um_equation_boquien}
L_{TIR}=216.9\times (\nu L_\nu)^{0.836}_{PAH 8\mu m} (\pm29\%) 
\end{eqnarray}
Note that for  \citet{2010ApJ...713..626B}, we used their conversion in Table 1 from SINGS data. The unit is also converted from W to $L_{\odot}$.
We overplot these relations in Fig.\ref{fig:relation}.
There is good agreement between these conversions and our best-fit relation, especially at  $9<log L_{TIR}<11$, where both samples have enough number of galaxies.
Note that \citet{2007ApJ...660...97C}'s sample spans at $0<z<0.6$ and \citet{2008A&A...479...83B}'s is mostly at $z\leq 0.4$. Thus, these samples are significantly at higher redshifts than ours. SED evolution could cause a small difference.
See  \citet{2010ApJ...713..626B} for metallicity dependence of the conversion.

For $L_{12\mu m}$, there exists a conversion from  \citet{2005A&A...432..423T}.
\begin{eqnarray}\label{eq:12um_takeuchi}
\log L_{TIR}=1.02+0.972 \log L_{12\mu m},
\end{eqnarray}
We overplot the relation in Fig.\ref{fig:relation}.
The slope is in good agreement. There is a $\sim$50\% offset in zero points. 
Possible reasons of this offsets include different definition in $L_{12\mu m}$ and different estimates for $L_{TIR}$.
 Our  $L_{12\mu m}$ relation is virtually identical to the one given by \citet{1995ApJ...453..616S}
which is, in our units:
\begin{eqnarray}\label{eq:12um_spinoglio}
\log L_{TIR}=1.51+0.942 \log L_{12\mu m},
\end{eqnarray}

For $L_{24\mu m}$, we compare the conversions with the following in the literature \citep{2010ApJ...713..626B,2008A&A...479...83B}.
\begin{eqnarray}\label{24um_equation_boquien}
L_{TIR}=57.9\times (\nu L_\nu)^{0.923}_{24\mu m} (\pm54\%) 
\end{eqnarray}
\begin{eqnarray}\label{24um_equation_bavouzet}
L_{TIR}=6856\times (\nu L_\nu)^{0.71}_{24\mu m} (\pm54\%) 
\end{eqnarray}
Our conversion is in very good agreement with  \citet{2010ApJ...713..626B}.  \citet{2008A&A...479...83B}'s relation has a slightly shallower slope, but in good agreement at $9<log L_{TIR}<11$, where most of their galaxies lie.

\citet{2007ApJ...656..770S} observed 17$\mu$m PAH complex features in Spitzer IRS spectra of nearby star-forming galaxies. 
This feature was not detected previously with ISO, and therefore not included in the SED model we used.
Although \citet{2007ApJ...656..770S}'s SED templates only contained two data points in the far-IR, and thus were not suitable for our purpose,
we try to estimate how the  17$\mu$m PAH complex features affect the  $L_{15\mu m}$-$L_{TIR}$ conversion as follows.
First, we remove a continuum from both of the spectra, \citep{2001ApJ...556..562C,2007ApJ...656..770S}.
 Then we replace the 6-20$\mu$m region of the \citet{2001ApJ...556..562C}'s spectra by \citet{2007ApJ...656..770S}.
 Here we scaled the spectrum so that the  amplitude of the 12$\mu$m PAH complex matches.
 Thus, this new templates keep the overall SED shape of \citet{2001ApJ...556..562C}, but have the 17$\mu$m PAH complex features from \citet{2007ApJ...656..770S}.
 We rerun the SED fit to estimate the  $L_{15\mu m}$-$L_{TIR}$ conversion to find that $L_{TIR}$ estimate (as a function of $L_{15\mu m}$) will be $\sim$3\% smaller if the  17$\mu$m PAH complex features are included in the SED. Although the effect is smaller than the errors of the conversion, this indicates that future SED models need to include the 17$\mu$m PAH complex features.



With current technology, detectors are much more sensitive in the mid-IR range than in the far-IR range.
Therefore, once these correlations are assumed, they will provide a useful conversion to compute luminosities of many more galaxies, either at higher redshift or at a fainter flux level, from a single MIR flux.
 For example, the above relations can provide a total IR flux measurement for all sources from the AKARI MIR all sky survey, which has 3 times more sources than the FIR all sky survey. WISE will perform even deeper all sky survey in 3.4, 4.6, 12 and 22 $\mu m$ in the near future \citep{2008EAS....33...57W}. For future use, we also provide  $L_{12\mu m}$-$L_{TIR}$, and  $L_{22\mu m}$-$L_{TIR}$ conversions in WISE filters as follows, and also in Fig.\ref{fig:relation}.

The best-fit relation between  $L_{WISE 12\mu m}$ and  $L_{TIR}$ is 
\begin{eqnarray}\label{eq:wise12um}
  L_{TIR} = (18\pm5) \times \nu L_{\nu, WISE 12\mu m}^{0.96\pm0.01} (\pm 23\%).
\end{eqnarray}

The best-fit relation between  $L_{WISE 22\mu m}$ and  $L_{TIR}$ is 
\begin{eqnarray}\label{eq:22um}
  L_{TIR} = (53\pm20) \times \nu L_{\nu,22\mu m}^{0.91\pm0.01} (\pm 25\%).
\end{eqnarray}

\section{Infrared Luminosity functions}
\subsection{The 1/$V_{\max}$ method}\label{sec:vmax}

With accurately measured $L_{TIR}$, we are ready to construct IR LFs.
Since the RBGS is a flux-limited survey, we need to correct for a volume effect to compute LFs.
We used the 1/$V_{\max}$ method \citep{1968ApJ...151..393S} for this. An advantage of the 1/$V_{\max}$  method is that it allows us to compute a LF directly from data, with no parameter dependence or a model assumption. A drawback is that it assumes a homogeneous galaxy distribution and thus is vulnerable to local over-/under-densities \citep{2000ApJS..129....1T}. 

A comoving volume associated to any source of a given luminosity is defined as $V_{\max}=V_{z_{\max}}-V_{z_{\min}}$, where $z_{\min}$ is the lower limit of the redshift  and $z_{\max}$ is the maximum redshift at which the object could be seen given the flux limit of the survey. 
 In this work, we set $z_{\min}$=0.0004 since at a very small redshift, an error in redshift measurement is dominated by a peculiar motion, and thus,  $L_{TIR}$ also has a large error. This only removes 5 galaxies from the sample.

For the RBGS, the detection limit is IRAS $S_{60\mu m}=$5.24Jy.
We used the SED templates 
 \citep{2001ApJ...556..562C}
for k-correction to obtain the maximum observable redshift from the flux limit.

 For each luminosity bin then, the LF is derived as

\begin{eqnarray}
\phi =\frac{1}{\Delta L}\sum_{i} \frac{1}{V_{\max,i}} \label{LF}
\end{eqnarray}

\noindent ,where $V_{\max}$  is a comoving volume over which the $i$th galaxy could be observed, and $\Delta L$ is the size of the luminosity bin (0.3 dex). 
The RBGS is a complete in 60$\mu$m at IRAS $S_{60\mu m}>$5.24Jy. Completeness correction in terms of sky coverage of both satellites is taken into account.

\subsection{Monte Carlo simulation}\label{sec:monte}
 Uncertainties in the LF values stem from various factors such as 
 the finite numbers of sources in each luminosity bin, 
 the k-correction uncertainties,
 and the flux errors. 
 To compute these errors we performed Monte Carlo simulations by creating 1000 simulated catalogs, where 
 each catalog contains the same number of sources, but we assign each source new fluxes following a Gaussian distribution centered at fluxes with a width of a measured error.
 Then we measured errors of each bin of the LF based from the variation in the 1000 simulations.
 These estimated errors are added in quadrature to the Poisson errors in each LF bin.

\begin{table*} 
 \centering
  \caption{Best double power-law fit parameters for the AKARI LFs, and the infrared luminosity density obtained from the fit}\label{tab:fit_parameters}
  \begin{tabular}{@{}cccccccccc@{}}
  \hline
Sample &  $L_{TIR}^*$ ($L_{\odot}$)&  $\phi^*(\mathrm{Mpc^{-3} dex^{-1}})$ & $\alpha$ (faint-end)& $\beta$ (bright-end) &$\Omega_{IR}(L_{\odot} Mpc^{-3})$ \\ 
 \hline
 \hline
Total & 7.8$\pm0.2\times 10^{10}$ &   0.00037$\pm$0.00005   & 1.8$\pm$0.1 &      3.6$\pm$0.1 & 8.5$^{+1.5}_{-2.3}\times 10^{7}$	\\
 \hline
SFG   &   10$\pm0.2\times 10^{10}$ &   0.00018$\pm$0.00006   & 1.9$\pm$0.1 &      4.4$\pm$0.3 &	 8.0$^{+1.7}_{-2.5}\times 10^{7}$\\
AGN   &  7.8$\pm0.2\times 10^{10}$ &   0.00011$\pm$0.00005   & 1.7$\pm$0.1 &      3.2$\pm$0.3 &  1.6$^{+0.5}_{-0.3}\times 10^{7}$	\\
\hline
\end{tabular}
\end{table*}

\subsection{IR luminosity function}\label{sec:LF}

\begin{figure}
\begin{center}
\includegraphics[scale=0.6]{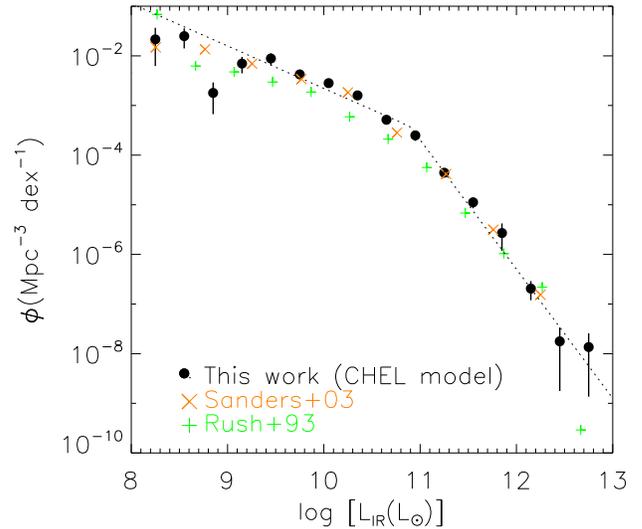}
\end{center}
\caption{
Infrared luminosity function of the RBGS. The $L_{TIR}$ is measured using the AKARI 9,18,65,90,140 and 160$\mu$m fluxes through an SED fit. Errors are computed using 1000 Monte Carlo simulations, added by Poisson error.
The dotted lines show the best-fit double-power law. 
The crosses show data from \citet{2003AJ....126.1607S}, who measured $L_{TIR}$  using IRAS photometry.
The plus show data from \citet{1993ApJS...89..1R}, who measured $L_{FIR}$ by integrating IRAS fluxes over 12-100$\mu$m.
}\label{fig:LF}
\end{figure}

\begin{figure}
\begin{center}
\includegraphics[scale=0.4]{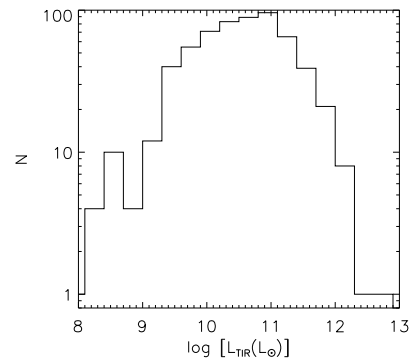}
\end{center}
\caption{A luminosity histogram of galaxies used to compute Fig.\ref{fig:LF}.
}\label{fig:Nhist}
\end{figure}

In Fig.\ref{fig:LF}, we show LF of the RBGS but using AKARI photometry. 
Fig.\ref{fig:Nhist} shows number of galaxies used to compute the LF.
The original LF measured using the IRAS photometry is also overplotted. There is a very good agreement between the LFs measured by AKARI and IRAS over the luminosity range of 8$<$log$L_{TIR}<$12. Although the IRAS-based  $L_{TIR}$ is computed using a single polynomial equation (Eq.\ref{eq:sanders}), the agreement shows it measured the IR LF very well. Perhaps, since a LF is an integrated quantity, it can be measured more reliably than $L_{TIR}$ of individual galaxies, even using data up to  only 100$\mu$m.

Our LF in Fig.\ref{fig:LF} agrees well with \citet{1993ApJS...89..1R}, after correcting their results from a Hubble constant of 50 to 75.
 Their LF slightly underestimate LF, perhaps because they only integrated IRAS photometry up to 100$\mu$m.
Our LF
also agrees well with \citet{2003ApJ...587L..89T}, once their 60$\mu$m LF is converted to $L_{TIR}$ by multiplying 2.5 \citep{2006A&A...448..525T}. Their LF was measured in a SED model-free way, using different density estimators.
It is reassuring that LFs measured in  completely different ways agree well.
See Sedgwick et al. (in prep.) for a similar attempt in the AKARI Deep field south.

Following \citet{2003AJ....126.1607S}, we fit an analytical function to the LFs.
In the literature, IR LFs were fit better by a double-power law \citep{2006MNRAS.370.1159B,Goto_NEP_LF} or a double-exponential \citep{1990MNRAS.242..318S,2004ApJ...609..122P,2006A&A...448..525T,2005ApJ...632..169L} than a Schechter function, which steeply declines at the high luminosity and underestimates the number of bright galaxies.  
In this work, we fit the TIR LFs using a double-power law \citep{2006MNRAS.370.1159B}.
\begin{equation}
 \label{eqn:lumfunc2p}
 \Phi(L)dL/L^{*} = \Phi^{*}\bigg(\frac {L}{L^{*}}\bigg)^{1-\alpha}dL/L^{*}, ~~~ (L<L^{*})
\end{equation}
\begin{equation}
 \label{eqn:lumfunc2p2}
 \Phi(L)dL/L^{*} = \Phi^{*}\bigg(\frac {L}{L^{*}}\bigg)^{1-\beta}dL/L^{*}, ~~~  (L>L^{*})
\end{equation}
\noindent Free parameters are $L^*$ (characteristic luminosity, $L_{\odot}$), $\phi^*$ (normalization, Mpc$^{-3}$), $\alpha$, and $\beta$ (faint and bright end slopes), respectively.
The best-fit values  are summarized in Table \ref{tab:fit_parameters}. The local LF has a break at $L^*$=7.8$\pm0.2\times 10^{10}$. Understanding how this break ($L^*$) evolves as function of cosmic time, and what causes the break is fundamental to galaxy evolution studies. This work provides an important benchmark in the local Universe.



\subsection{Bolometric IR luminosity density based on the TIR LF}\label{sec:omega_IR}
One of the primary purposes in computing IR LFs is to estimate the IR luminosity density,  which in turn is a good estimator of the dust-hidden cosmic star formation density \citep{1998ARA&A..36..189K}, provided the AGN contribution is removed. 
 The bolometric IR luminosity of a galaxy is produced by thermal emission of its interstellar matter. In SF galaxies, the UV radiation produced by young stars heats the interstellar dust, and the reprocessed light is emitted in the IR. For this reason, in star-forming galaxies (SFG), the bolometric IR luminosity is a good estimator of the current SFR (star formation rate) of the galaxy.

Once we measured the LF,  we can estimate the total infrared luminosity density by integrating the LF, weighted by the luminosity. We used the best-fit double-power law to integrate outside the luminosity range in which we have data, to obtain estimates of the total infrared luminosity density, $\Omega_{TIR}$.

The resulting total luminosity density is  $\Omega_{TIR}$= (8.5$^{+1.5}_{-2.3})\times 10^{7}$ $L_{\odot}$Mpc$^{-3}$.
Errors are estimated by varying the fit within 1$\sigma$ of uncertainty in LFs.
 Out of  $\Omega_{TIR}$, 7$\pm1$\% is produced by LIRG ($L_{TIR}>10^{11}L_{\odot}$), and only 0.4$\pm$0.1\% is by  ULIRG ($L_{TIR}>10^{12}L_{\odot}$). A very small fraction of  $\Omega_{TIR}$ is produced by luminous infrared galaxies at z=0.0082, in stark contrast to high-redshift Universe. 
We found that $\sim30$\% of $\Omega_{TIR}$ originates from $L_{TIR}<10^{8.2}L_{\odot}$, where we do not have data and had to rely on the extrapolation of the faint-end tail of the LF. Therefore, similar amount of uncertainty cannot be ruled out if the faint-end slope changes significantly under $L_{TIR}<10^{8.2}L_{\odot}$.
We will discuss the evolution of  $\Omega_{TIR}$ in Section \ref{sec:evolution}.



\section{Infrared luminosity density and its evolution}
\subsection{Separating IR contributions from AGN and SFG}\label{sec:f_agn}


We showed a local IR LF in Fig.\ref{fig:LF}. However this includes both IR emission from star formation activity and from AGN. Since different physics govern star formation and AGN, it is essential to separate their IR emission to understand the cosmic evolution of each component. 
For example, the cosmic star formation history cannot be addressed without subtracting the contribution from the AGN, and vise versa.
 
 Separating AGN from starbursts is not a trivial task.  Many AGN/SF separation methods have been proposed such as X-ray, radio luminosity, optical line ratios, PAH strengths, submm properties and so on, many of which often disagree with each other. 
We do not have a complete diagnosis of AGN/SB such as emission line ratios and X-ray for all of our sample. 
Therefore instead of classifying individual galaxies into AGN and SF,
 we attempt a statistical approach to separate contributions to LFs by AGN and star-forming galaxies.
 
\begin{figure}
\begin{center}
\includegraphics[scale=0.6]{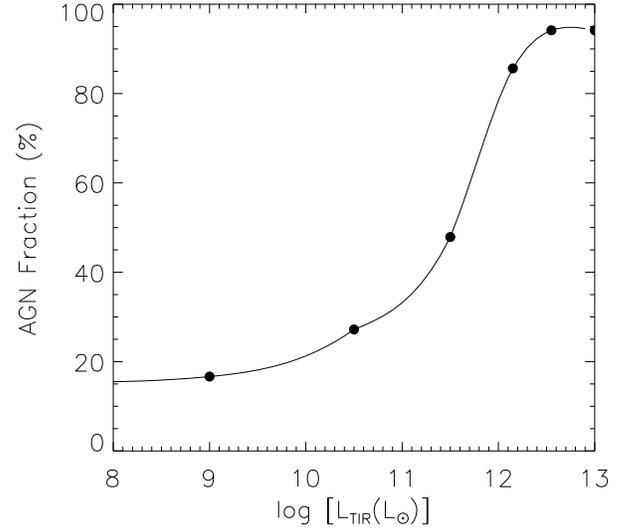}
\end{center}
\caption{
Fractions of AGN is shown as a function of $L_{TIR}$. 
The original classification was taken from Fig.4 of \citet{2010ApJ...709..884Y}, and interpolated by a spline curve to be applied to our LFs. 
}\label{fig:f_AGN}
\end{figure}

 In this work, we simply use a result of a new classification of \citet{2010ApJ...709..884Y}, which used three carefully examined optical line ratio diagrams  \citep{2006MNRAS.372..961K} to classify individual galaxies.   Based on optical line ratios they classify IR galaxies into AGN (LINER, Seyfert 1 and 2), star-forming, and composite as a function of $L_{TIR}$ in their Table 4 and Figure 4. Their results agree well with other work \citep{2003ApJ...587L..89T,2005MNRAS.360..322G}, within 20\% or so.
 We use their fractions of AGN/SFG as a function of $L_{TIR}$, to separate our LF into that of AGN and SFG.
 In this process, it is an open question how to handle composite galaxies, which are close to 50\% of galaxies in almost all bins.  In this work, we simply assigned the AGN/SF fractional ratio of that luminosity bin to separate composite galaxies (and ambiguous galaxies) into AGN/SF galaxies. 
 We understand this is not an ideal approach. There will be up to $\sim$50\% uncertainty if these composite galaxies are all AGN or SFG.
 It merely provides one method to separate IR LFs into those of AGN and SFG. 
 A method to assess IR contribtuion by AGN and SFG is becoming realistic. For example, \citet{2009A&A...502..457G} showed a tight correlation between nuclear $L_{12.3\mu m}$ and  $L_{X-ray}$.
 In the near future, more realistic classification of composite galaxies will become possible.
We show the resulting AGN fraction from this process as a function of $L_{TIR}$ in Fig.\ref{fig:f_AGN}. 
At $L_{TIR}>10^{12}L_{\odot}$, this procedure assigns 90\% of IR luminosity to AGN. 
At $L_{TIR}<10^{10}L_{\odot}$, the AGN fraction is 20\% or less.
This also agrees well with a recent measurement by \citet{2010ApJ...709..572K}.

\begin{figure}
\begin{center}
\includegraphics[scale=0.6]{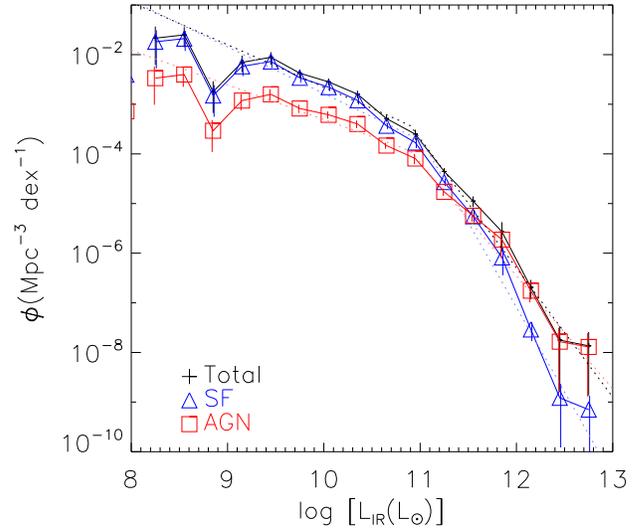}
\end{center}
\caption{
IR LF is separated for star-forming galaxies (blue triangles) and AGN (red squares) using Fig.\ref{fig:f_AGN}. Total IR LF is shown with   the black plus sign. The dotted-lines are the best-fit double-power laws.
}\label{fig:AGN_LF}
\end{figure}

By applying this AGN/SF separation, in Fig.\ref{fig:AGN_LF} we show LFs separately for AGN and SFG. 
The total IR LF shown in  Fig.\ref{fig:LF} is also shown for comparison.
As expected from  Fig.\ref{fig:f_AGN}, 
 Fig.\ref{fig:AGN_LF} shows that at  $L_{TIR}>10^{12}L_{\odot}$, the AGN are responsible for almost all of the $L_{TIR}$, forming a very steep bright-end drop for the SFG LF.
On the other hand, at $L_{TIR}<10^{11}L_{\odot}$, SF galaxies explain most of the  $L_{TIR}$.

 Next, we fit a double-power law (Eqs. \ref{eqn:lumfunc2p} and \ref{eqn:lumfunc2p2}) to the AGN and SF IR LFs, exactly as we did for the total IR LF in Fig.\ref{fig:LF}.
 The best-fit parameters are summarized in Table \ref{tab:fit_parameters}.
 The most notable difference is at the bright-end slope ($\beta$), where the AGN LF has a very shallow slope of $\beta$=3.2$\pm$0.3, while the  SFG has a steep slope of $\beta=$4.4$\pm$0.3.
As expected, the faint-end tail ($\alpha$) is not much different, in fact, being consistent with each other within 1$\sigma$.
The $L^*$ of SFG becomes brighter due to the steepening of the bright-end slope.

The total infrared luminosities of AGN and SFG's were also measured in separated LF's by  \citet{1993ApJS...89..1R}.
Adding their Seyfert 1 and Seyfert 2 LF's, and correcting for different Hubble constant, they get somewhat lower
AGN/SFG LF ratios than our overall value of 1.6 : 8.0.
Their methodology was simpler, and did not account for the population of LINER AGN in their 12$\mu$m-selected galaxy sample.


%


What do these differences in LFs bring to the IR luminosity density, $\Omega_{IR}$, by AGN and SFG?
We estimate  the total infrared luminosity density by integrating the LFs weighted by the luminosity, separately for AGN and SFG. We used the double power law outside the luminosity range in which we have data, to obtain estimates of the total infrared luminosity density, $\Omega_{TIR}$, for AGN and SFG.

\begin{table}
 \centering
  \caption{Local IR luminosity densities
}\label{tab:omega}
  \begin{tabular}{@{}lrrlllcccc@{}}
  \hline
  Sample & $\Omega_{IR}^{SFG}$ ($L_{\odot}$Mpc$^{-3}$)  & $\Omega_{IR}^{AGN}$ ($L_{\odot}$Mpc$^{-3}$)    \\
 \hline
 \hline
Total    & (8.0$^{+1.7}_{-2.5})\times 10^{7}$   & (1.6$^{+0.5}_{-0.3})\times 10^{7}$  \\
LIRG    &  (3.8$^{+0.4}_{-0.9})\times 10^{6}$   & (2.1$^{+0.6}_{-0.5})\times 10^{6}$  \\
ULIRG     & (1.5$^{+1.0}_{-1.0})\times 10^{4}$ & (12$^{+5}_{-7})\times 10^{4}$ \\
 \hline
\end{tabular}
\end{table}

The resulting total luminosity density ($\Omega_{IR}$) is,
$\Omega_{IR}^{SFG}$=8.0$^{+1.7}_{-2.5}\times 10^{7}$ $L_{\odot}$Mpc$^{-3}$, and
$\Omega_{IR}^{AGN}$= 1.6$^{+0.5}_{-0.3}\times 10^{7}$ $L_{\odot}$Mpc$^{-3}$, as summarized in Table \ref{tab:omega}.
Errors are estimated by varying the fit within 1$\sigma$ of uncertainty in LFs.
 These are also summarized in Table \ref{tab:fit_parameters}.
The results show that among the total IR luminosity density integrated over all the IR luminosity range,
  83\% ($\frac{\Omega_{IR}^{SFG}}{\Omega_{IR}^{AGN}+\Omega_{IR}^{SFG}}$) of IR luminosity density is emitted by the SFG, and only 16\% ($\frac{\Omega_{IR}^{AGN}}{\Omega_{IR}^{AGN}+\Omega_{IR}^{SFG}}$) is by AGN at z=0.0082. 
 The results shows that at low redshift (z=0.0082), majority of the IR luminosity density is emitted by the SFG. Therefore, even if you convert all the $\Omega_{IR}$ into SFR density, you will only overestimate by 20\%.
 However, the situation is different at higher redshift, where a number of papers reported luminosity evolution in the IR LF \citep{2005ApJ...630...82P,2005ApJ...632..169L,2009A&A...496...57M}. The AGN/SFG separation will have much larger effect and thus more important at higher redshift.

Once we have  $\Omega_{IR}^{SFG}$, we can estimate star formation density emitted in infrared light.
 The SFR and $L_{TIR}$ is related by the following equation for a Salpeter IMF, 
$\phi$ (m) 
$\propto m^{-2.35}$ between 
$0.1-100 M_{\odot}$  \citep{1998ARA&A..36..189K}.
\begin{eqnarray}
SFR [M_{\odot} yr^{-1}] =1.72 \times 10^{-10} L_{TIR} [L_{\odot}] 
\end{eqnarray}
By using this equation, we obtain SFR density = 1.3$\pm$0.2 $10^{-2}M_{\odot} yr^{-1}$.

If we limit our integration to ULIRG luminosity range  ($L_{TIR}>10^{12}L_{\odot}$), we obtain,
    $\Omega_{IR}^{SFG}$(ULIRG)= (1.5$^{+1.0}_{-1.0})\times 10^{4}$ $L_{\odot}$Mpc$^{-3}$, and
    $\Omega_{IR}^{AGN}$(ULIRG)= (12$^{+5}_{-7})\times 10^{4}$ $L_{\odot}$Mpc$^{-3}$.
In other words, at ULIRG luminosity range, AGN explain 88\%  ($\frac{\Omega_{IR}^{AGN}(ULIRG)}{\Omega_{IR}^{AGN}(ULIRG)+\Omega_{IR}^{SFG}(ULIRG)}$) of IR luminosity, again showing the AGN dominance at the bright-end.
 
In the LIRG luminosity ($L_{TIR}>10^{11}L_{\odot}$), results are
    $\Omega_{IR}^{SFG}$(LIRG)= (3.8$^{+0.4}_{-0.9})\times 10^{6}$ $L_{\odot}$Mpc$^{-3}$, and
    $\Omega_{IR}^{AGN}$(LIRG)= (2.1$^{+0.6}_{-0.5})\times 10^{6}$ $L_{\odot}$Mpc$^{-3}$.
This shows  AGN contribution is already down to 35\%  ($\frac{\Omega_{IR}^{AGN}(LIRG)}{\Omega_{IR}^{AGN}(ULIRG)+\Omega_{IR}^{SFG}(LIRG)}$) of IR luminosity in the LIRG range.

%
%
%

\subsection{Evolution of $\Omega_{IR}^{SFG}$}\label{sec:evolution}

\begin{figure*}
\begin{center}
\includegraphics[scale=1.1]{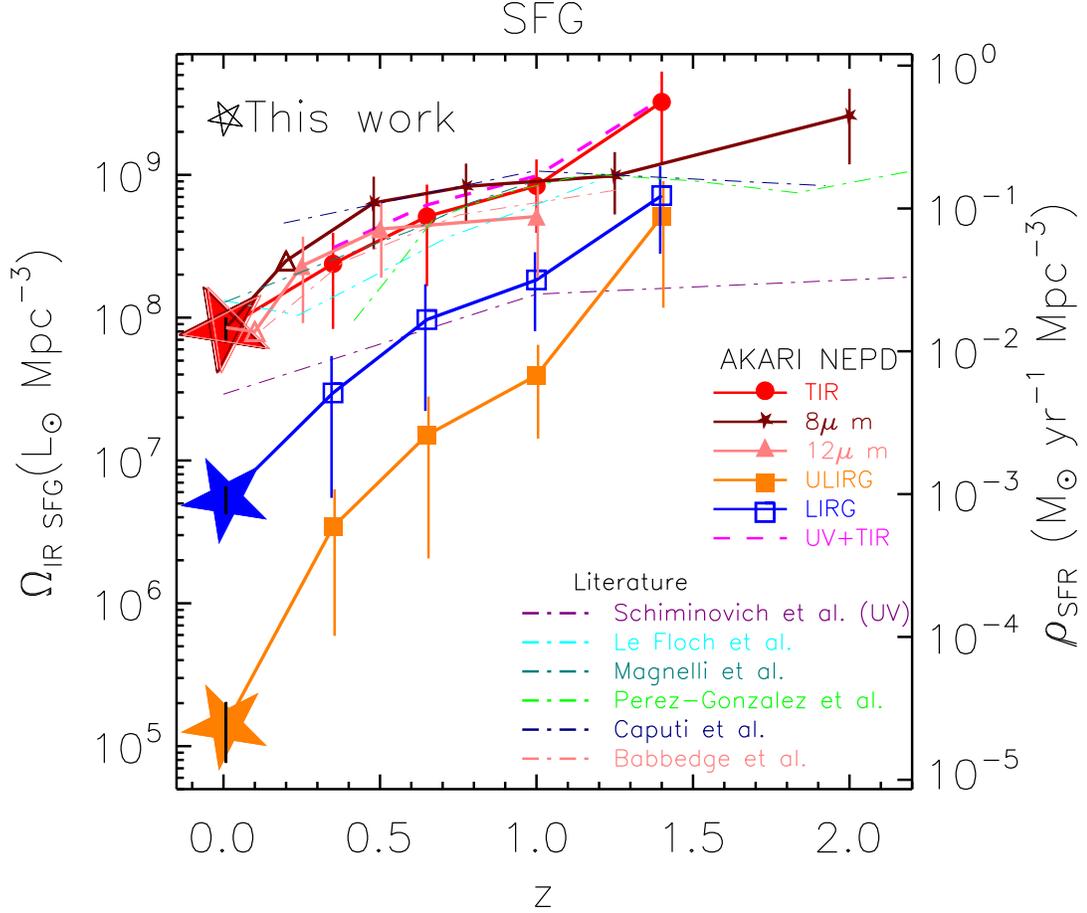}
\end{center}
\caption{
Evolution of TIR luminosity density by star-forming galaxies. Results from this work is plotted with triangles at z=0.0082. The red, blue and orange triangles show IR luminosity density from all galaxies, from LIRG only, and from ULIRG only. 
Higher redshift results in the solid lines are from the AKARI NEP deep field \citep{Goto_NEP_LF}.
 Shown with different colors are TIR luminosity density based on TIR LFs (red circles), 8$\mu$m LFs (stars), and 12$\mu$m LFs (filled triangles). The blue open squares and orange filled squares  are for only LIRG and ULIRGs, also based on our $L_{TIR}$ LFs.
Overplotted dot-dashed lines are estimates from the literature: \citet{2005ApJ...632..169L}, \citet{2009A&A...496...57M}, \citet{2005ApJ...630...82P}, \citet{2007ApJ...660...97C},   and \citet{2006MNRAS.370.1159B} are in cyan, yellow, green, navy, and pink, respectively.
The purple dash-dotted line shows UV estimate by \citet{2005ApJ...619L..47S}.
The pink dashed line shows the total estimate of IR (TIR LF) and UV \citep{2005ApJ...619L..47S}.
}\label{fig:TLD_all}
\end{figure*}

We have separated the  $\Omega_{IR}^{SFG}$ from  $\Omega_{IR}^{AGN}$. 
Now we are ready to examine the evolution of $\Omega_{SFR}$ without contribution from AGN.
In Fig.\ref{fig:TLD_all}, we plot the evolution of  $\Omega_{IR}^{SFR}$ as a function of redshift.
Higher redshift results are taken from \citet{Goto_NEP_LF}, who also tried to exclude AGN using SED fitting to individual galaxies.
 Results from the Spitzer survey and GALEX survey are also plotted.
The   $\Omega_{IR}^{SFG}$ shows a strong evolution as a function of redshift. The best-fit linear relation is 
  $\Omega_{IR}^{SFG}\propto$(1+z)$^{4.0\pm0.5}$.
This is consistent with most of earlier work. For example, 
\citet{2005ApJ...632..169L} obtained $\gamma = 3.9\pm 0.4$ up to z$\sim$1.
\citet{2005ApJ...630...82P} found $\gamma = 4.0\pm 0.2$ from z=0 to 0.8.
\citet{2006MNRAS.370.1159B} obtained $\gamma = 4.5^{+0.7}_{-0.6}$.
\citet{2009A&A...496...57M} obtained $\gamma = 3.6\pm 0.4$ up to z=1.3.
\citet{2010A&A...515A...8R} found $\gamma = 3.8\pm 0.4$ in the redshift interval of $0<z<1$.
\citet{2010A&A...518L..27G} found $\gamma = 3.8\pm 0.3$ up to z$\sim$1, with some evidence of flattening up to z$\sim$2.

Once the IR luminosity density is separated into ULIRG and LIRG contribution,
we found
  $\Omega_{IR}^{SFG} (ULIRG)\propto$(1+z)$^{9.1\pm0.8}$, and
  $\Omega_{IR}^{SFG}(LIRG)\propto$(1+z)$^{5.3\pm2.0}$.
 $\Omega_{IR}^{SFG}(ULIRG)$ shows more rapid evolution than $\Omega_{IR}^{SFG} (LIRG)$, showing importance of luminous IR sources at high redshift.

\subsection{Evolution of $\Omega_{IR}^{AGN}$}\label{sec:agn_evolution}

In turn, we can also investigate the evolution of $\Omega_{IR}^{AGN}$. 
This has been difficult in the literature since it is difficult to separate faint AGN individually in the presence of a host galaxy.
By integrating IR LF$_{AGN}$ in Fig.\ref{fig:AGN_LF}, we obtained $\Omega_{IR}^{AGN}$=1.6$^{+0.5}_{-0.3}\times 10^{7}$ $L_{\odot}$Mpc$^{-3}$. 

   \citet{2010ApJ...709..572K} recently reported that the mean SED of 70$\mu$m-selected sample from the COSMOS survey at $z\sim$1 is similar to what has been observed locally. As a natural consequence, the fraction of AGNs as a function of $L_{IR}$ is in excellent agreement with that in the local universe (their Fig.30). 
If so, we can apply the $f_{AGN}$-$L_{IR}$ relation described in Fig.\ref{fig:f_AGN} to IR LFs at higher redshifts to investigate the evolution of $\Omega_{IR}^{AGN}$.

Practically, we applied the $f_{AGN}$-$L_{IR}$ relation in Fig.\ref{fig:f_AGN} to IR LFs at $0<z<1.5$ presented by \citet{Goto_NEP_LF} to obtain $\Omega_{IR}^{AGN}$.
 \citet{Goto_NEP_LF} used their own AGN/star-forming galaxy classification to remove AGN from their LFs, but for a fair comparion, we applied the same methodology as used in Section \ref{sec:f_agn} to their LFs before they subtract the contribuion from individual AGN.

\begin{figure*}
\begin{center}
\includegraphics[scale=.7]{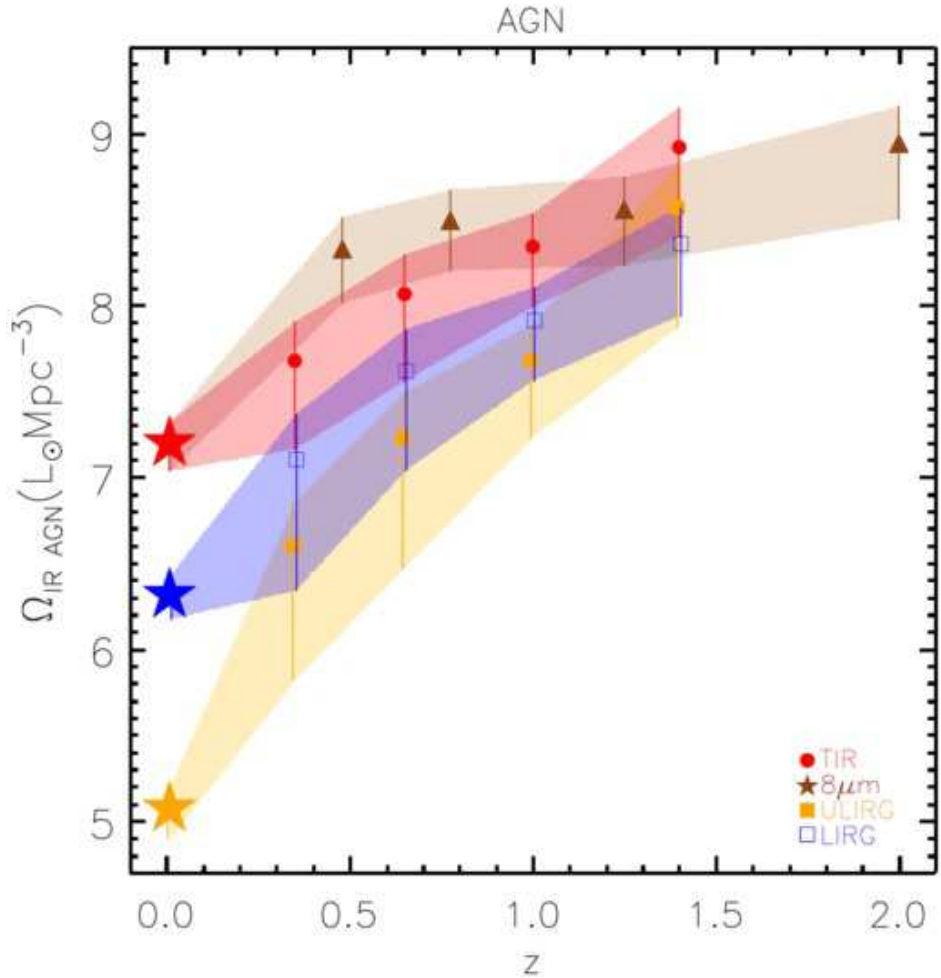}
\end{center}
\caption{
Evolution of TIR luminosity density by AGN.
 Results from this work is plotted with stars at z=0.0082. The red, blue and orange points show IR luminosity density from all AGN, from LIRG AGN only, and from ULIRG AGN only. 
Higher redshift results are from the AKARI NEP deep field \citep{Goto_NEP_LF}, with contribution from star forming galaxies removed.
 Brown triangles are $\Omega_{IR}^{AGN}$ computed from the 8$\mu$m LFs \citep{Goto_NEP_LF}.
}\label{fig:TLD_AGN_all}
\end{figure*}

In Fig.\ref{fig:TLD_AGN_all}, we show the evolution of $\Omega_{IR}^{AGN}$, which shows a strong evolution with increasing redshift. 
 At a first glance, both $\Omega_{IR}^{AGN}$ and $\Omega_{IR}^{SFG}$ show rapid evolution, suggesting that the correlation between star formation and black hole accretion rate continues to hold at higher redshifts, i.e., galaxies and black holes seem to be evolving hand in hand.
However, there are some differences as well.
 When we fit the evolution with (1+z)$^{\gamma}$, we found 
 $\Omega_{IR}^{AGN}\propto$(1+z)$^{4.4\pm0.4}$, which is possibly more rapid evolution than  $\Omega_{IR}^{SFG}$ although errors are significant.
The contribution by ULIRGs quickly increases toward higher redshift;  By z=1.5, it exceeds that from LIRGs. Indeed, we found $\Omega_{IR}^{AGN}(ULIRG)\propto$(1+z)$^{9.4\pm0.8}$ and $\Omega_{IR}^{AGN}(LIRG)\propto$(1+z)$^{5.5\pm0.8}$.

\begin{figure}
\begin{center}
\includegraphics[scale=0.6]{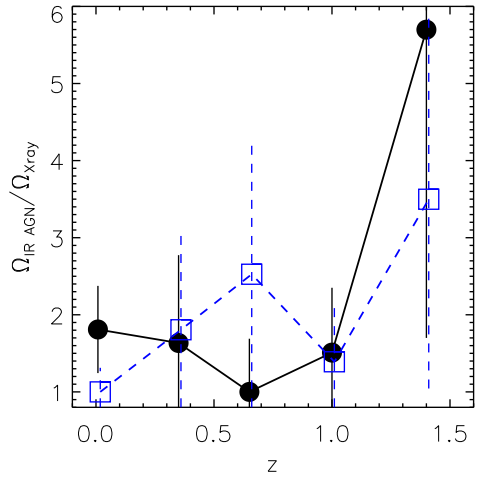}
\end{center}
\caption{
X-ray luminosity density to $\Omega_{IR}^{AGN}$ ratio is plotted as a function of redshift. 
The ratio is normalized at the minimum value. 
The X-ray luminosity densities of black circles and blue squares are from \citet{2010MNRAS.401.2531A} (2-10 keV) 
 and \citet{2005A&A...441..417H} (0.5-2 keV)   , respectively.
}\label{fig:IR_X_ratio}
\end{figure}

It seems there is no sign of flattening at high redshift in  $\Omega_{IR}^{AGN}(z)$. It would be interesting to compare this with a number-density evolution of optical QSOs \citep{2004MNRAS.349.1397C,2006AJ....131.2766R}, which peaks at z=2-3, and the evolution of X-ray AGN \citep{2003ApJ...598..886U,2005A&A...441..417H}. \citet{2010MNRAS.401.2531A}  recently presented the evolution of X-ray luminosity density, which shows a turnover at around z$\sim$1.2$\pm0.1$. In Fig.\ref{fig:TLD_AGN_all}, our  $\Omega_{IR}^{AGN}$ does not show any sign of decline at least up to z=1.5, in contrast to the X-ray results.
\citet{2005AJ....129..578B}  also showed much shallower evolution in X-ray than our IR results.
In Fig.\ref{fig:IR_X_ratio}, we investigate the evolution of X-ray luminosity density to  $\Omega_{IR}^{AGN}$ ratio. X-ray luminosity densities are taken from \citet{2010MNRAS.401.2531A} \citet{2010MNRAS.401.2531A} (2-10 keV)  and \citet{2005A&A...441..417H} (0.5-2 keV). 
It is interesting that the ratio is consistent with a constant value at $0<z<1$, and shows a possible increase at $z>1$, although it is unfortunate our errors are too large to draw a firm conclusion. 
The possible difference may suggest an increase of obscured AGN light toward higher redshift at $z>1$ compared with optical/X-ray unobscured AGN. Post-starburst AGN galaxies found by \citet{2006MNRAS.369.1765G} may be an example of such phenomena.
Recently, \citet{TreisterScience} showed an increase in the number ratio of obscured/unobscured QSOs with increasing redshift, concluding that a merger-driven black hole evolution model is consistent with the observed result.
 Our  Fig.\ref{fig:IR_X_ratio} may suggest  a similar obscured/unobscured AGN evolution in terms of AGN luminosity density, instead of the number density.
A possible deviation from $\Omega_{IR}^{SFG}$ also suggests AGN may have formed earlier than star-formation in terms of infrared luminosity density.
More accurate comparison of X-ray and IR luminosity density evolution of AGN will bring an interesting physical implication on the subject. 
In particular, reducing measurement errors at high redshift is urgent.



%


%
%
%
%
%
%
%

\section{Summary}

Using AKARI's 6-band IR photometry in 9, 18, 65, 90, 140, and 160$\mu m$ with much improved spatial resolution to resolve background cirrus/confusing sources, we have re-measured  $L_{TIR}$ of the RBGS through IR SED model fitting. This is a significant progress considering that the IRAS used one linear equation to estimate $L_{TIR}$, with only data at $<100\mu$m. 

By using this new $L_{TIR}$ measurement, we constructed local IR LFs separately for SFG and AGN.
We also computed local infrared luminosity density through the derived LFs, and compared $\Omega_{IR}^{SFG}$ and $\Omega_{IR}^{AGN}$ to those at higher redshifts.

Our findings are as follows.
\begin{itemize}
 \item SED model-to-model variation in estimating $L_{TIR}$ is less than $25\%$.
 \item  We present $L_{MIR}$-to-$L_{TIR}$ conversions for  Spitzer $8\mu m$, AKARI $9\mu$m, IRAS $12\mu m$, WISE $12\mu m$, ISO $15\mu m$, AKARI $18\mu m$, WISE $22\mu m$, and Spitzer $24\mu$m  filters. These conversions provides us with a useful tool to estimate $L_{TIR}$ with a MIR band only. 
 \item Re-constructed local TIR LF with the AKARI data is consistent with that from the IRAS, i.e., the AKARI's better data show the IRAS measurement was correct. 
 \item By integrating the IR LF weighted by  $L_{TIR}$, we obtain  the local cosmic IR luminosity density of $\Omega_{TIR}$= (8.5$^{+1.5}_{-2.3})\times 10^{7}$ $L_{\odot}$Mpc$^{-3}$.
 \item LIRG and ULIRG contribute to $\Omega_{TIR}$ a little;
Only  7$\pm1$\% of $\Omega_{TIR}$ is produced by LIRG ($L_{TIR}>10^{11}L_{\odot}$), and only 0.4$\pm$0.1\% is by  ULIRG ($L_{TIR}>10^{12}L_{\odot}$) in the local Universe, in stark contrast to high redshift results.
 \item  Compared with high redshift results from the AKARI NEP deep survey, we observed a strong evolution of $\Omega_{TIR}^{SFG}\propto$(1+z)$^{4.0\pm0.5}$, after removing AGN contribution.
 \item We showed an evolution of  $\Omega_{IR}^{AGN}$ scales as  $\propto$(1+z)$^{4.4\pm0.4}$. ULIRG contribution to the  $\Omega_{IR}^{AGN}$ exceeds that by LIRG by z=1.5.
 \item $\Omega_{IR}^{AGN}$/$\Omega_{X-ray}^{AGN}$ ratio shows a possible increase at $z>1$. If confirmed, this may suggest an increase of obscured AGN at $z>1$.
\end{itemize}

\section*{Acknowledgments}

We thank the anonymous referee for many insightful comments, which significantly improved the paper.
We are indebted to M.Malkan for many valuable comments and suggestions.
We thank D.Sanders and J.M. Mazzarella for useful discussions. 


T.G. acknowledges financial support from the Japan Society for the Promotion of Science (JSPS) through JSPS Research Fellowships for Young Scientists.



%

This research is based on the observations with AKARI, a JAXA project with the participation of ESA.

The authors wish to recognize and acknowledge the very significant cultural role and reverence that the summit of Mauna Kea has always had within the indigenous Hawaiian community.  We are most fortunate to have the opportunity to conduct observations from this sacred mountain.

Support for the work of ET was provided by the National Aeronautics and
Space Administration through Chandra Postdoctoral Fellowship Award
Number PF8-90055 issued by the Chandra X-ray Observatory Center, which
is operated by the Smithsonian Astrophysical Observatory for and on
behalf of the National Aeronautics Space Administration under contract
NAS8-03060.

TTT has been supported by Program for Improvement of Research
Environment for Young Researchers from Special Coordination Funds for
Promoting Science and Technology, and the Grant-in-Aid for the Scientific
Research Fund (20740105) commissioned by the Ministry of Education,
Culture,
Sports, Science and Technology (MEXT) of Japan.
TTT has been also partially supported from the Grand-in-Aid for the Global
COE Program ``Quest for Fundamental Principles in the Universe: from
Particles to the Solar System and the Cosmos'' from the MEXT.

%
%





\label{lastpage}

\end{document}